\documentclass[12pt]{amsart}
\usepackage{geometry}
\usepackage{graphicx}
\usepackage{dcolumn}
\usepackage{bm}
\usepackage{amsmath,amsfonts}
\usepackage{latexsym}%
\usepackage{lineno}   
\usepackage{array}
\usepackage{adjustbox}
\usepackage{natbib,url}

\newcommand{\ud}{\,\mathrm{d}}
\newcommand{\RR}{\mathbb{R}}

\begin{document}

\title[]{Analyzing transient-evoked otoacoustic emissions by concentration
of frequency and time}

\author{Hau-Tieng Wu}
\address{Hau-Tieng Wu\\
Department of Mathematics and Department of Statistical Science\\
Duke University, NC, 27705, USA}

\author{Yi-Wen Liu}
\address{Yi-Wen Liu\\
Department of Electrical Engineering\\
National Tsing Hua University, Hsinchu 30013, Taiwan}
\email{ywliu@ee.nthu.edu.tw}

\begin{abstract}
The linear part of transient evoked (TE) otoacoustic emission (OAE) is thought to be generated
via coherent reflection near the characteristic place of constituent wave components. Because of the tonotopic
organization of the cochlea,
high frequency emissions return earlier than low frequencies; however, due to the random
nature of coherent reflection, 
the instantaneous frequency (IF) and amplitude envelope of TEOAEs both fluctuate. Multiple reflection components and synchronized spontaneous emissions can further make it difficult to extract the IF by linear transforms.
In this paper, we propose to model TEOAEs as a sum of {\em intrinsic mode-type functions} and analyze it by a {nonlinear-type time-frequency analysis} technique called concentration of
frequency and time (ConceFT). When tested with synthetic OAE signals {with possibly multiple oscillatory components}, the present method
is able to produce clearly visualized traces of individual components on the time-frequency plane. Further,
when the signal is noisy, the proposed method is
 compared with existing linear and bilinear methods in its accuracy for estimating the fluctuating IF.  
 Results suggest that ConceFT outperforms the best of these methods
 in terms of optimal transport distance, reducing the error by 10 to {21\%} when the signal
 to noise ratio is 10 dB or below.
\end{abstract}

\maketitle

\section{INTRODUCTION}

\begin{table}[t]
\caption{List of abbreviations}
\begin{center}
\begin{tabular}{ll}
\hline
{\bf AM}  & Amplitude modulation \\
{\bf BM}   &  Basilar membrane \\
{\bf ConceFT}    & Concentration of frequency and time \\
{\bf CWD}  &   Choi-Williams distribution \\
{\bf CWT}  &  Continuous wavelet transform \\
{\bf IF}   &  Instantaneous frequency \\
{\bf EMD}  &  empirical mode decomposition \\
{\bf IMT}   &  Intrinsic mode type \\
{\bf iTFR}   & Ideal time-frequency representation \\ 
{\bf MT}   &  Multi-taper \\
{\bf RM}   &  Reassignment method \\
{\bf OAE}   &  Otoacoustic emission \\
{\bf OTD}   & Optimal transport distance \\ 
{\bf SNR}   & Signal-to-noise ratio \\
{\bf SFOAE}   &  Stimulus frequency OAE \\
{\bf SOAE}   &  Spontaneous OAE \\
{\bf SPWV}    & Smoothed pseudo Wigner-Ville distribution \\
{\bf SSOAE}  &   Synchronized spontaneous OAE \\
{\bf SST}   &  Synchrosqueezing transform \\
{\bf STFT}    & Short-time Fourier transform \\
{\bf TB}    & Tone burst \\
{\bf TEOAE}   &  Transient evoked OAE \\
{\bf T-F}   & Time-frequency \\
\hline
\end{tabular}
\end{center}
\label{tab:1}
\end{table}%
{Transient evoked (TE) otoacoustic emissions (TEOAEs)} were discovered 40 years ago \cite{Kemp-1978}. By inspecting
the waveforms, TEOAEs generally exhibit a \emph{chirp-like} feature in that the high frequency components seem to occur earlier than low frequency parts. 
The latency of TEOAE as a function of frequency (hereafter referred to as the \emph{latency function}) could potentially
provide valuable information for hearing diagnostic purposes because it is, if not directly proportional to, 
at least highly correlated with the sharpness of cochlear and psychoacoustic tuning \citep{NeelyEA-1988, SheraEA02, SheraEA-2010}. 
However,  
{ the notion of latency function itself might be an over-simplification of the signal of interest for a few reasons.}
First, the coherent reflection theory predicts that 
the phase in TEOAE {spectra} would fluctuate \cite{ZweigShera-1995, TalmadgeEA-2000} because of the random nature
in how the traveling waves are scattered in the cochlea near the characteristic places. 
The spectrum varies across different ears, and the  
latency function unavoidably has large deviations if directly derived from the phase gradient \citep{SheraBergevin-2012}. 
Secondly, the reverse traveling waves in the cochlea also get reflected at the stapes, so the 
OAE waveform measured in the ear canal is a superposition of multiple reflections { \cite[e.g., see Fig.~8 of][]{TalmadgeEA-1998}}. 
{ To capture the reflections, we propose that TEOAE signals would be represented better by
a sum of {{\em intrinsic mode type (IMT) functions}} that each has time-varying amplitudes modulation
and instantaneous frequency.} Based on {this} model, we suggest to analyze TEOAE waveforms
by a modern {time-frequency (T-F) analysis} tool called \emph{concentration of frequency and time} (ConceFT) \cite{Daubechies-EA-2016}.

Modern {T-F analysis tools} could be roughly classified into three categories, the linear-type, the bilinear-type and the nonlinear-type. The basic ideas behind the linear-type T-F analysis include (i) dividing the signal into segments and evaluating the spectrum for each segment (e.g., the Gabor transform or short-time Fourier transform (STFT)) \cite{Flandrin:1999}, (ii) measuring the similarity between the signal and a series of dilations of a given mother wavelet, which leads to the continuous wavelet transform (CWT) \cite{Tognola1997,Notaro2007}, {or (iii) the S-transform (ST) \cite{MISHRA2016161} that combines features from STFT and CWT such as frequency modulation and dilation}. The bilinear-type T-F analysis catches signal properties from the energy or cross correlation viewpoint, which includes a wide range of methods from the traditional Wigner-Ville distribution to the Cohen class \cite{Flandrin:1999}. The nonlinear-type T-F analysis aims to depict the signal in a more data-driven way, by either taking more signal information to modify the linear-type or bilinear-type T-F analyses, or extracting the information directly from the signal. This category includes the reassignment method (RM) \cite{Auger_Flandrin:1995}, the synchrosqueezing transform (SST) \cite{Daubechies_Lu_Wu:2011, Wu:2011Thesis,Oberlin2015}, the empirical mode decomposition (EMD) \cite{Huang_Shen_Long_Wu_Shih_Zheng_Yen_Tung_Liu:1998,KopsinisMcLaughlin-2009}, and several others. We refer the readers to \cite{Flandrin:1999} for a general discussion of available methods and \cite{Daubechies-EA-2016} for a recent review of the field.

Specifically for analyzing TEOAE, 
\cite{JedrzejczakEA-2009} built a dictionary of asymmetric Gabor functions to span a linear space and applied matching pursuit algorithms to identify the best fit to {TEOAEs}. The latency function could be inferred and empirical fits were reported. The {CWT} 
has been utilized to see the composition or frequency variation of the TEOAE \cite{Tognola1997,Notaro2007}, to filter the OAE signals \cite{Janusauskas2001,Moleti2012}, to investigate the relationship of TEOAE latency and stimulus level \cite{SistoMoleti-2007}, and
to infer the hearing functionality of neonates  
\citep{MoletiEA-2005,Tognola2001}. Because of the multi-resolution property, CWT provides flexibility to {analyze the frequency latency structure} of the OAEs. 
By filtering on the T-F plane and then applying the inverse transform, it became also possible to separate the first-reflection component from its mixture with later reflections \cite{SheraBergevin-2012}.
Recently, the robustness against noise for various time-frequency techniques was compared using simulated TEOAE data \citep{BiswalMishra-2017},  the techniques being compared included STFT,  CWT,  SST, the S-transform, and EMD;
the result suggested that CWT was the most accurate way for estimating the latency function at a signal-to-noise ratio (SNR) of 15 dB.

As elegant as the existing T-F analysis techniques are, however, there are intrinsic difficulties toward a deeper insight into the TEOAE. For the widely applied linear-type T-F analysis, like STFT, CWT {and S-transform}, the uncertainty principle \cite{Flandrin:1999,Ricaud2014} is inevitable. 
A direct consequence of the uncertainty principle is a blurring of the spectrum, depending on the chosen window and its length. 
Another limitation is its dependence on the chosen window or mother wavelet and lack of the adaptivity to the signal. {For example, while we can take the frequency latency property of TEOAE signal into account and design a mother wavelet to well track that part \cite{Tognola1997, SistoEA2015}, when there are other components in the recording, such as the synchronized
spontaneous OAE (SSOAE) \cite[e.g., discussed by][]{Keefe-2012}, we might need another mother wavelet to catch them. In short, how to choose a universal mother wavelet to accomodate signals of different features is in general challenging for the linear method, and this could also} challenge the interpretation of the outcomes.

While the bilinear-type T-F analysis has a potential to alleviate these limitations, it suffers from different limitations. For example, while the adaptivity issue is resolved by taking the signal itself as the window in the traditional Wigner-Ville distribution, it is limited by the interference terms when the signal is composed of multiple oscillatory components or time-varying frequency. For other commonly applied Cohen class algorithms, like Choi-Williams distribution (CWD) or Smoothed pseudo Wigner-Ville distribution (SPWV), the same interference issue persists. {When choosing windows for smoothing is needed, the same interpretation issue for the linear method still stands. Furthermore, when time information is touched during smoothing, it is difficult to preserve causality}.

Among different nonlinear-type T-F analyses, the widely applied EMD lacks of theoretical foundation and might lead to erroneous interpretation and conclusion for real data. RM and SST, on the other hand, have been developed rigorously with theoretical support to handle the traditional T-F analysis limitations. In particular,  by taking the phase information of the signal into account, the spectrum is sharpened beyond the blurriness limit caused by the uncertain principle, and the resulting T-F representation is less dependent on the chosen windows \cite{Daubechies_Lu_Wu:2011, Wu:2011Thesis,Oberlin2015}. {Depending on how the phase information is used, SST can be classified into 
first \cite{Wu:2011Thesis} or second \cite{Oberlin2015} order. The first order SST is limited to signals with slowly varying frequency and the second order SST is designed to handle fast varying frequency situation.} {While it is a nonlinear method, it is shown in \cite{Chen_Cheng_Wu:2014} that the first order SST is robust to {reasonable amount of} different types of noise, including the non-stationary and heteroscedastic kinds}. However, when the signal to noise ratio (SNR) is low (e.g., below 2 dB), the nonlinear-type T-F analyses {in general does} not perform well.
To sum up, while there have been several T-F analysis techniques, it is still a long lasting challenge to study the TEOAE signal, due to (i) its intrinsic oscillatory structure predicted by theory, (ii) the existence of multiple reflection components or even synchronized spontaneous emissions \cite{Keefe-2012}, and (iii) the low SNR encountered in practice.

To handle these challenges, in this paper, we explore the possibility of analyzing TEOAEs by ConceFT \cite{Daubechies-EA-2016}, which is a nonlinear-type T-F technique that extends the RM and SST by combining the \emph{multi-tapering {(MT)}} technique. 
It has been established that, if the signal of interest can be modeled as a sum of {IMT functions satisfying} a \emph{well-separated} condition and certain \emph{slow-varying} assumptions, then ConceFT helps produce sharpened traces on the T-F plane that represent the signal and are robust to noise \cite{Daubechies-EA-2016}.
The basic idea behind ConceFT is twofold. First, {a nonlinear-type T-F analysis is chosen, like SST or RM, and the} sharpened time-frequency representation { provides higher fidelity to the spectral content of the signal.}
Second, the effects of noise in OAE measurement are reduced by {generalizing the traditional MT} technique \cite{Percival:1993,Daubechies-EA-2016}, which benefits directly from the nonlinearity of the sharpening procedure. 

To understand which kind of information ConceFT could accurately extract, we {adopt} two different models to simulate
TEOAE in a controlled manner, so that we can evaluate the performance of ConceFT analysis thoroughly and understand
what are the conditions for it to work well on TEOAEs. 
In one model, the TEOAE spectrum is
expressed in terms of a direct integral that takes the presence of irregularity into account \cite{SheraBergevin-2012, ZweigShera-1995}. In the other model \cite{LiuNeely-2010}, irregularities are present in physical variables such as the basilar-membrane mass, damping coefficients, and stiffness, and TEOAEs are ``measured'' by time-domain simulation. {Lastly, we quantify the performance and compare ConceFT with other T-F analysis tools on a fully simulated signal that we know the ground truth.}
 
The organization of the rest of this paper is as follows. Sec.~\ref{sec:new2} introduces SST and ConceFT in details. Sec.~\ref{sec:new3} shows the results of applying SST and ConceFT to analyze synthetic TEOAE data. 
Based on the results of simulation, Sec.~\ref{sec:new4} discusses the effectiveness
and limitation of the proposed signal analysis approach. Conclusions are given in Sec.~\ref{sec:6}.

\section{Concentration of frequency and time}\label{sec:new2}

Based on the literature review, we understand that a TEOAE signal may contain multiple reflection components or even SSOAEs. We suggest that it would be appropriate to model each component with 
time-varying amplitude and frequency. 
To capture these, we 
{resort} to the intrinsic {mode} type (IMT) function \cite{Daubechies_Lu_Wu:2011} defined as follows, 
\[
f(t)= A(t)e^{i \phi(t)}, 
\]
where  $A(t)$ and $\phi(t)$ satisfy the following three conditions. 
The first condition is the \textit{regularity} condition; that is, $A$ and $\phi$ are smooth enough. 
The second one is the \textit{boundedness} condition; that is, both $A(t)$ and $\phi'(t)$ are strictly positive and bounded from above.
The third one is the \textit{slowly varying} condition; that is, we could find 
{constants $\varepsilon_1,\varepsilon_2>0$} so that 
{$|A'(t)|\leq\varepsilon_1 |\phi'(t)|$ and $|\phi''(t)|\leq\varepsilon_2|\phi'(t)|$ for all $t$.}\footnote{ Note that this definition is slightly different from that given in \cite{Daubechies_Lu_Wu:2011}, where $\varepsilon_1=\varepsilon_2$, since for a practical signal, like TEOAE, the physical units of $A(t)$ and $\phi'(t)$ might be different. This unit issue could be taken care by $\varepsilon_1$ and $\varepsilon_2$.}
We refer to $A(t)$ as the {\em amplitude modulation} (AM) or amplitude envelope of $f(t)$, $\phi(t)$ the phase function of $f(t)$, and $\phi'(t)$ the {\em (angular) instantaneous frequency} (IF) of $f(t)$. The regularity, boundedness and slowly-varying conditions say that locally an IMT function behaves like a { sinusoidal} function. {The slowly varying {IF} condition can be slightly relaxed \cite{Oberlin2015,KOWALSKI201889} to accommodate fast varying {IF}
 like chirps, but to simplify the discussion, we focus on the slowly varying conditions.}  
Note that this model satisfies the identifiability condition \cite{Chen_Cheng_Wu:2014}; that is, if we could find $a(t)$ and $\psi(t)$ that satisfy $f(t)= A(t)e^{i \phi(t)}=a(t)e^{i \psi(t)}$, then $a(t)=A(t)$, and $\psi(t)=\phi(t)$ up to a global difference of an integer multiple {of $2\pi$} for all time $t$. This can be easily seen by taking the absolute value and hence unwrapping the phase.
Note that although the IMT function is written in the complex form, it is in general not analytic, 
{ since the Fourier transform of an IMT function might not be supported on the positive axis due to the time-varying amplitude and frequency.}

We thus model the TEOAE signal as a sum of several IMF functions, with different AM and IF functions. For instance, when only two components are considered, the adaptive harmonic model has the following expression, 
\begin{equation}
f(t) = A_1(t)e^{i\phi_1(t)}+A_2(t)e^{i\phi_2(t)},
\end{equation}
where $A_1(t)e^{i\phi_1(t)}$ may model the dominant first-reflection component of TEOAE and $A_2(t)e^{i\phi_2(t)}$ models the second reflection. {In this setup, we need the {\em frequency separation} condition; that is, $\phi_1'(t)-\phi_2'(t)\geq d>0$ \cite{Wu:2011Thesis}, or $\phi_1'(t)-\phi_2'(t)\geq d(\phi'_1(t)+\phi'_2(t))$ \cite{Daubechies_Lu_Wu:2011}. Different linear T-F analysis tool needs different frequency separation properties, depending on the {frequency} modulation or dilation nature of the T-F analysis; { $\phi_1'(t)-\phi_2'(t)\geq d>0$ is needed if STFT is considered, and $\phi_1'(t)-\phi_2'(t)\geq d(\phi'_1(t)+\phi'_2(t))$ is needed if CWT is considered.} The frequency separation condition is needed for the identifiability condition to be satisfied \cite{Chen_Cheng_Wu:2014,Oberlin2015,KOWALSKI201889}.}

{ If a TEOAE signal can be modeled as a sum of IMT functions}, several modern nonlinear-type T-F analysis tools can be applied with theoretical guarantees. Particularly, we can apply the RM or SST to analyze the TEOAE signal and expect to get the IF and AM information back \cite{Daubechies_Lu_Wu:2011}.\footnote{In reality, TEOAE signals are real-valued, and Hilbert transform
has been applied to construct an IMT from its real part \cite{Keefe-2012}. In contrast, the RM or SST methods do
not require Hilbert transform to take place because they
can handle concurrents IMFs, including the special case of 
$\cos\phi(t) = \frac{1}{2}(e^{i\phi(t)}+e^{-i\phi(t)})$.}
Further, when the SNR is low, we could {consider} the MT technique \cite{Percival:1993,Xiao_Flandrin:2007} to stabilize the noise impact. {A combination of {nonlinear-type T-F analysis tools, like SST,} and MT techniques is called \emph{concentration of frequency and time} (ConceFT).} The SST and the MT techniques are described next, respectively.

\subsection{Synchrosqueezing transform}
The basic idea behind SST is taking the phase information hidden inside the chosen T-F representation, like STFT \cite{Wu:2011Thesis}, CWT \cite{Daubechies_Lu_Wu:2011} {or S-transform \cite{7229315}}, and shuffling the T-F representation coefficients to alleviate the {blurring effect} caused by the uncertainty principle. Specifically, SST is composed of three steps. {Below, we discuss SST based on STFT, and the discussion for CWT {or S-transform} can be found in \cite{Daubechies_Lu_Wu:2011,7229315}.}
First, for a given properly defined function $f$, 
the STFT associated with a window function $h(t)$ is defined by
\begin{equation}\label{Definition:STFT}
V_f^{(h)}(t,\nu):=\int f(\tau)h(\tau-t)e^{-i2\pi \nu (\tau-t)}\ud \tau\,,
\end{equation}
where $t\in\RR$ is the time, $\nu = \omega/2\pi \in \RR^+$ is the frequency, $h$ is the window function chosen by the user --- a common choice is the Gaussian function, i.e. $h(t)=(2\pi\sigma)^{-1/2}e^{-t^2/{2}\sigma^2}$, where $\sigma >0$. 
To sharpen the spectrogram $|V_f^{(h)}(t,\nu)|^2$, note that the phase information of $V_f^{(h)}(t,\nu)$ is not used. A keen observation made in \cite{Kodera_Gendrin_Villedary:1978,Auger_Flandrin:1995,Flandrin:1999} is that the geometric and phase information in the T-F representation allows us to sharpen it. 

To motivate this keen observation, consider a simple function $f(t)=Ae^{i 2\pi f_0 t}$, where $A\,,f_0>0$. By a direct calculation, $V_f^{(h)}(t,\nu)=A\hat{h}(\nu-f_0)e^{i 2\pi f_0 t}$, where $\hat{h} := \mathcal{F}[h(t)]$. {Note that the frequency $f_0$ shows up in the phase of $V_f^{(h)}(t,\nu)$. A na\"{i}ve idea to obtain the frequency information thus consists of two steps: first, the partial derivative of $V_f^{(h)}(t,\nu)$ associated with $t$ is calculated, which gives $\partial_t V_f^{(h)}(t,\nu)=i2\pi f_0A\hat{h}(\nu-f_0)e^{i 2\pi f_0 t}$, and
then the frequency $f_0$ can be retrieved by a direct division; that is, 
\begin{equation}
f_0 = \frac{\partial_t V_f^{(h)}(t,\nu)}{i2\pi V_f^{(h)}(t,\nu)}. 
\end{equation}
To avoid calculating the numerical derivative $\partial_t V_f^{(h)}(t,\nu)$, note that 
\begin{equation}
\partial_t V_f^{(h)}(t,\nu)=-V_f^{(\mathcal D h)}(t,\nu)+i2\pi\nu V_f^{(h)}(t,\nu),
\end{equation}
where $\mathcal{D}h$ denotes the derivative of $h$ with respect to time. We thus have $V_f^{(\mathcal D h)}(t,\nu)=i2\pi(\nu-f_0) V_f^{(h)}(t,\nu)=i2\pi A(\nu-f_0)\hat{h}(\nu-f_0)e^{i 2\pi f_0 t}$. This observation motivates the definition of the following {\em reassignment rule} \cite[Definition 2.3.12]{Wu:2011Thesis}}, 
\begin{equation}
\omega^{(h)}_{f}(t,\nu):=
\nu - \mathfrak{Im}\frac{V_f^{(\mathcal{D}h)}(t,\nu)}{2\pi V_f^{(h)}(t,\nu)},
\label{eq:reass}
\end{equation}
where $\mathfrak{Im}$ means taking the imaginary part. Equation~(\ref{eq:reass}) is well-defined on every points $(t,\nu)$ where $V_f^{(h)}(t,\nu)\neq 0$.

{
To sharpen the T-F representation of $V^{(h)}_f(t,\nu)$, an intuitive approach is, for each time $t$, moving all coefficients to the entry associated with the frequency we have interest. This intuition is carried out in SST by the following integration formula \cite[Definition 2.3.13]{Wu:2011Thesis}:}
\begin{equation}\label{Definition:1stSST}
s^{(h)}_{f}(t,\nu):=
\int_{\mathfrak{N}_t}V^{(h)}_f(t,\nu')\delta_{|\nu-\omega^{(h)}_f(t,\nu')|}\ud \nu',
\end{equation}
where $\mathfrak{N}_t:=\{\nu':\,|V^{(h)}_f(t,\nu')|>0\}$. 
{ 
Equation (\ref{Definition:1stSST}) can be understood as a combination of two steps:
\begin{itemize}
\item{
First, find all entries $(t,\nu')$ so that the frequency information provided by $\omega^{(h)}_f(t,\nu')$ is $\nu$, which is embodied in $\delta_{|\nu-\omega^{(h)}_f(t,\nu')|}$.\footnote{To make it mathematically rigorous, the delta measure should be replaced by a smooth approximation. We skip the technical detail here for the simplicity.}} 
\item{
Secondly, gather all non-zero STFT coefficients to the entry $(t,\nu)$ by the integration.} 
\end{itemize}
In the simple example $f(t)=Ae^{i 2\pi f_0 t}$, the non-zero STFT coefficients are all moved to $f_0$ in $s^{(h)}_{f}(t,\nu)$, resulting a sharp T-F representation.
}

As is shown in \cite[Theorem 2.3.14]{Wu:2011Thesis}, SST can be applied to study IMT functions. For $f(t)=A(t)e^{i\phi(t)}$, the sharpened spectrogram by SST (that is, $|s^{(h)}_{f}(t,\nu)|^2$) is concentrated on $\phi'(t)/2\pi$ with the AM function $A(t)$ encoded inside. {As a result, this technique alleviates the blurring effect caused by the uncertainty principle.}

{However, when the IF of an IMT function changes rapidly,
the outcome of the above-mentioned SST 
becomes less ideal because the reassignment rule is again ``blurred''. Certain improvement has been found by \cite{Oberlin2015} via further manipulation of the phase function to accommodate the fast-varying IF;} the following reassignment rule was introduced,
\begin{align}
\Omega^{(h)}_{f}(t,\nu)&=
\left\{
\begin{array}{ll}
\omega^{(h)}_{f}(t,\nu)+Q^{(h)}_f(t,\nu)(t-T^{(h)}_f(t,\nu))&\mbox{when }  {\partial_\nu} T^{(h)}_f(t,\nu)\neq 0\\
\omega^{(h)}_{f}(t,\nu) & \mbox{otherwise},\end{array}\right.\label{RM:omega}
\end{align}
where
\begin{align}
&Q^{(h)}_f(t,\nu):=\frac{V_f^{(\mathcal{D}\mathcal{D} h)(t,\nu)}(t,\nu)V_f^{(h)}(t,\nu)-(V_f^{(\mathcal{D} h)}(t,\nu))^2 }{2\pi i[(V_f^{(h)}(t,\nu))^2+V_f^{(\mathcal{T}h)}(t,\nu)V_f^{(\mathcal{D}h)}(t,\nu)-V_f^{(\mathcal{T}\mathcal{D}h)}(t,\nu)V_f^{(h)}(t,\nu)]} \nonumber\\
&T^{(h)}_f(t,\nu):=t+\mathfrak{Re}\frac{V_f^{(\mathcal{T} h)}(t,\nu)}{V_f^{(h)}(t,\nu)} \label{RM:omega} ,
\end{align}
$Q_f$ and $T_f$ are defined when their denominators are not zero, $\mathfrak{Re}$ means taking the real part, and  $(\mathcal{T} h)(t):=th(t)$.
$\Omega^{(h)}_{f}$ is called the {\em second-order frequency reassignment rule} \cite{Oberlin2015}. 
{
With this terminology, we may call $\omega^{(h)}_{f}$ the {\em first-order frequency reassignment rule} \cite{Wu:2011Thesis}.}

With the second-order frequency reassignment rule, the second order STFT-based SST is defined as: 
\begin{equation}\label{definition:SST}
S^{(h)}_{f}(t,\nu):=
\int_{\mathfrak{N}_t}V^{(h)}_f(t,\nu')\delta_{|\nu-\Omega^{(h)}_f(t,\nu')|}\ud \nu'.
\end{equation}
{Again, we may call $s^{(h)}_{f}(t,\nu)$ the first-order STFT-based SST.
Note that,  for both first and second order STFT-based SST,} we nonlinearly reassign the STFT coefficient {\em only} on the frequency axis, so the causality of the signal is preserved and hence the reconstruction is possible, although we do not pursue these properties in this work.  Yet another property of first-order SST is its robustness to noise of different kinds \cite{Brevdo_Fuckar_Thakur_Wu:2013,Chen_Cheng_Wu:2014}, while it is nonlinear in nature.  
The above properties {enable} us to {extract} dynamical information of a noisy oscillatory signal, particularly the IF and AM.

In this work, due to the chirp-like behavior of the TEOAE signal, we consider the second-order STFT-based SST for the analysis. With no danger of confusion, we call it SST for simplicity below, {unless we specify that it is the first order SST}. {Note that CWT-based SST can also be considered \cite{Daubechies_Lu_Wu:2011} for the analysis, but to simplify the discussion, we focus only on the STFT-based SST.}

\subsection{Generalized multi-taper}

While {the first order} SST is {theoretically shown to be} robust to a mild level of noise \cite{Brevdo_Fuckar_Thakur_Wu:2013,Chen_Cheng_Wu:2014}, when the noise is large, its performance might be jeopardized. {While theoretical analysis for other nonlinear-type T-F analysis tools are not available, empirically they are robust only when noise level is mild.} In practice, the TEOAE 
obtained within a limited amount of time could be noisy with a rather small SNR. 
Thus, a technique to reduce the effect of noise would be desired. 
In this work, we consider the recently proposed generalized MT technique \cite{Daubechies-EA-2016} to achieve this goal. 

{%
The spirit of the traditional MT technique roots in the law of large numbers. Ideally, from a recorded noisy signal, if we can generate several copies of information composed of the clean component and independent noise, by taking average the clean signal will be enhanced. This intuitive idea is carried out in the following way.} With the chosen orthonormal windows, the obtained spectral information associated with the clean signal is almost invariant among windows, while that associated with the noise is independent. 
{%
For example, take a noisy signal given as $Y=f+\xi$, where $f$ is the signal we have interest and $\xi$ is the added noise, and take $J$ orthonormal windows $h_j$, $j=1,\ldots,J$. By the linearity of STFT, we have $V^{(h_j)}_Y(t,\eta)=V^{(h_j)}_f(t,\eta)+V^{(h_j)}_\xi(t,\eta)$, for $i=1,\ldots, J$.  When $\xi$ is Gaussian and white, we see that  $V^{(h_j)}_\xi(t,\eta)$ and $V^{(h_k)}_\xi(t,\eta)$ are independent when $j\neq k$. While the spectral information associated with $V^{(h_j)}_f(t,\eta)$ depends on $h_j$, by the linearity of STFT again, $\frac{1}{J}\sum_{j=1}^J V^{(h_j)}_f(t,\eta)=V^{(\frac{1}{J}\sum_{j=1}^Jh_j)}_f(t,\eta)$.} 
Therefore, by taking an average, only the spectral information associated with the clean signal is reserved.  
The T-F representation of SST can be improved by the MT technique by considering
\begin{equation}
M_Y:=\frac{1}{J}\sum_{j=1}^JS_Y^{(h_j)}.
\end{equation}
In \cite{Lin_Wu_Tsao_Yien_Hseu:2014}, the MT technique combined with SST is applied to study the anesthesia depth. The combination of the MT technique and RM is considered in \cite{Xiao_Flandrin:2007}.
Ideally, if there are infinitely many orthonormal functions ``well supported'' in time and frequency domains, the MT technique would lead us to a low bias and low variance estimator of the clean signal information. 
However, it has been well studied in \cite{Daubechies:1988} that the number of orthonormal functions that are well concentrated in the T-F plane is limited. This fact is understood as the ``Nyquist rate'' for the 
T-F analysis.\footnote{{ The ``Nyquist rate'' here is different from the common Nyquist rate encountered in the sampling theory. Here, it is called the Nyquist rate to describe limited possible windows under the constraints such as orthonormality and well concentration. See \cite{Daubechies:1988} for a full development and \cite[ESM-4]{Daubechies-EA-2016} for a summary.}}

{To conquer the limitation of Nyquist rate and further stabilize the algorithm, the nonlinear nature of SST is considered. How ConceFT generalizes the above-mentioned} traditional MT technique could be manifested {by directly showing the algorithm. Consider a linear combination of given $J$ orthonormal windows $h_j$, $j=1,\ldots,J$,} $h=r^*H$, where $H=[h_1,\ldots,h_J]^T$, 
$r = [r_1, r_2, ..., r_J]\in \mathbb{C}^J$ and $|r|=1$, we could obtain a T-F representation, denoted as $S_Y^{(h)}$, by applying the SST. Note that when $r = e_j$ is the unit vector with the $j$-th entry equal to $1$, $r^*H=h_j$. 
The T-F representation based on ConceFT is defined as
\begin{equation}
C_Y:=\frac{1}{N}\sum_{n=1}^NS_Y^{(h_{(n)})}
\label{eq:C_y}
\end{equation}
where $r_{(n)}$ is uniformly sampled from the unit sphere in $\mathbb{C}^J$ and $h_{(n)}:=r_{(n)}^*H$.

Intuitively, due to the nonlinear nature of SST, the { level of dependence is reduced between noise components coming from non-orthogonal windows}. 
Thus, by taking average, the noise could be further canceled. To appreciate the importance of the nonlinearity, take the following examples into account. If we consider the linear-type T-F analysis, like the STFT, and 
{ follow the above-mentioned argument regarding generalized MT}, we have 
\begin{equation}
\lim_{N\to\infty}\frac{1}{N}\sum_{n=1}^NV_Y^{(h_{(n)})}=\lim_{N\to\infty}\frac{1}{N}\sum_{n=1}^N\sum_{j=1}^Je_j^\top r_{(n)}V_Y^{(h_j)}=\sum_{j=1}^J \big[\lim_{N\to\infty}\frac{1}{N}\sum_{n=1}^Ne_j^\top r_{(n)}\big]V_Y^{(h_j)} =0
\end{equation} 
due to the linearity of STFT. 
Here we use the fact that 
\[\lim_{N\to\infty}\frac{1}{N}\sum_{n=1}^N e_j^\top r_{(n)}=e_j^\top\int_{x\in\mathbb{C}^J;\,\|x\|=1}xdx=0.
\] 
Further, if we apply the generalized MT to the spectrogram, we have 
\begin{align*}
&\lim_{N\to\infty}\frac{1}{N}\sum_{n=1}^N|V_Y^{(h_{(n)})}|^2=\lim_{N\to\infty}\frac{1}{N}\sum_{n=1}^N|\sum_{j=1}^Je_j^\top r_{(n)}V_Y^{(h_j)}|^2\\
=\,& \sum_{j=1}^J \big[\lim_{N\to\infty}\frac{1}{N}\sum_{n=1}^N|e_j^\top r_{(n)}|^2\big]|V_Y^{(h_j)}|^2+\sum_{j\neq k}^J\big [\lim_{N\to\infty}\frac{1}{N}\sum_{n=1}^N(e_j^\top r_{(n)})^*e_k^\top r_{(n)}\big]{V_Y^{(h_j)}}^*V_Y^{(h_k)}\\
=\,& c\frac{1}{J}\sum_{j =1 }^J |V_Y^{(h_j)}|^2\,,
\end{align*}
for some constant $c>0$, where the last equality comes from the fact that 
\[ \lim_{N\to\infty}\frac{1}{N}\sum_{n=1}^N |e_j^\top r_{(n)}|^2=\int_{x\in\mathbb{C}^J;\,\|x\|=1}|x_j|^2 dx=c
\] and $\lim_{N\to\infty}\frac{1}{N}\sum_{n=1}^N(e_j^\top r_{(n)})^*e_k^\top r_{(n)}=\int_{x\in\mathbb{C}^J;\,\|x\|=1}x_j^*x_k dx=0$ due to the symmetry since $j\neq k$. Thus, the generalized MT technique leads to the traditional MT on the spectrogram. The same discussion holds for CWT and scalogram.

However, the situation is different when we apply the generalized MT technique to any nonlinear-type T-F analysis. For example, due to the nonlinearity of the SST,  $\lim_{N\to \infty}C_Y$ is not proportional to $M_Y$, since $S_Y^{(h_{(n)})}\neq  \sum_{j=1}^Je_j^\top r_{(n)}S_Y^{(h_j)}$. 
{In brief, due to the Nyquist rate limitation, we choose a reasonably small $J$, and count on a large $N$ to reject the noise.}
Although a complete theoretical quantification is still under study, a partial {theoretical result and numerical evidence in \cite{Daubechies-EA-2016}} show that { the level of dependence between noise components} caused by two non-orthonormal windows is { reduced} after SST. As a result, $\lim_{N\to \infty}C_Y$ is much closer to $S_f$ than $M_Y$ is, when measured by the optimal transportation distance \cite{Daubechies-EA-2016}. {We also mention that the generalized MT technique could be applied to other nonlinear-type T-F analysis, like RM. }

In practice, we choose $h_1,\ldots,h_J$ to be the first $J$ orthonormalized Hermite functions {because they are the most concentrated windows in the T-F domain \cite{Daubechies:1988,Daubechies-EA-2016}}. In particular, $h_1$ is the Gaussian function. In practice, $J$ could be chosen as small as $2$, while $N$ could be chosen as large as the user wishes, but a number of $N=30$ to $100$ is in general good enough \cite[e.g.][]{Daubechies-EA-2016}.

{
\section{Comparison of various ways to analyze TEOAE data}\label{sec:new3}
In this section, we report results of analyzing simulated TEOAE data by synchrosqueezing and ConceFT. The performance will be 
compared against what can be achieved by linear analysis methods, including STFT and CWT, and bilinear methods like CWD and SPWV. 
}

\subsection{Direct simulation of coherent reflection}\label{sec:cohref}
This section follows \citeauthor{ZweigShera-1995}'s \emph{coherent reflection} approach to simulate a TEOAE signal 
(\citeyear{ZweigShera-1995}).
Essentially, the TEOAE is regarded as the summation of reflected waves 
caused by mechanical irregularities throughout the entire length of the cochlea, which is treated
as a one-dimensional waveguide. Via Wentzel-Kramers-Brillouin (WKB) approximation \cite{Schroeder-1973}, \citeauthor{ZweigShera-1995} showed that the amount of reflection can be computed by introducing the irregularities to the wave propagation 
equation as a perturbation term. 
This framework successfully explained the periodic fine structure in the spectrum of different types of OAEs
{ and the hearing threshold}
\cite{TalmadgeEA-1998}. Here, we 
borrow a phenomenological equation that stems from the \citeyear{ZweigShera-1995} work for \emph{synthesizing} a TEOAE signal
without discussing the details of micromechanics of the cochlea. The reflection of traveling waves  in the cochlea due to unknown irregularities $\epsilon(x)$ is
written as follows \citep{TalmadgeEA-2000, SheraBergevin-2012, BiswalMishra-2017},
\begin{equation}
	R(\omega) = \int \epsilon(x') e^{\frac{-(x'-x_p)^2}{2(\Delta x)^2}}e^{-i4\pi\frac{x'-x_p}{\Lambda}}dx',
\label{eq:1}
\end{equation}
{where $\omega>0$}, $\Delta x$ represents a spatial spread of a traveling wave
near its characteristic place, and $\Lambda$ denotes the local wavelength.
Here, $R(\omega)$ represents a reflectance spectrum ``seen'' from the stapes ($x=0$) into the cochlea. In the equation, $x_p$ denotes the \emph{characteristic place} of (angular) frequency $\omega$ and is assumed to decrease against $\omega$ in a log-linear way as follows \citep{Greenwood-1990},
\begin{equation}
x_p(\omega) = l \log\frac{\omega_0}{\omega}, \label{eq:x_p}
\end{equation}
where $l$ is treated as a constant here,
and $\omega_0$ is the characteristic frequency at $x=0$. In Eq.~(\ref{eq:1}), the factor $\exp\left({\frac{-(x'-x_p)^2}{2(\Delta x)^2}}\right)$ describes the relative gain a small wave component receives
by going from $x=0$ to $x=x'$ and reflected to travel back to $x=0$, and the factor $\exp\{-i4\pi(x'-x_p)/\Lambda\}$ corresponds to the phase-shift thereof. In Eq.~(\ref{eq:1}), $R(\omega)$ is
zero if $\epsilon(x) = 0$, meaning that TEOAE would not exist if the cochlea were to be perfectly smooth.  

{
By fitting experimental data, it has been found that  $\Delta x$ and $\Lambda$ should
both vary slowly against $\omega$ \citep{SheraBergevin-2012}. 
Here, however, we set both of them as constants; in other words, we made
a crude simplification of cochlear mechanics so the model has 
a global scaling symmetry.

Also, we regard the inverse Fourier transform\footnote{
The inverse Fourier transform
was calculated via a 4096-point inverse fast Fourier transform with a sampling rate of 32 kHz.} 
$r(t) = \mathcal{F}^{-1}[R(\omega)](t)$ 
as a impulse response, which can be convolved with any transient acoustic stimulus to
calculate the first-reflection OAE evoked by the stimulus.\footnote{Of course, this would no longer be valid
if nonlinearity in cochlear mechanics is considered.} Thus, we ignore
multiple internal reflection in the cochlea due to impedance mismatch at the stapes. { The frequency response of reverse middle-ear transmission, the ear-canal acoustics, and the acoustic properties of the probe termination are also ignored, too.}

The model in Sec.~\ref{sec:cochmod} will consider both the multiple reflections in the cochlea and the middle-ear transmission, and restrict scaling symmetry to be valid 
only locally. Here in Sec.~\ref{sec:cohref}, 
the reason for making these crude simplifications are two-fold. First,
we shall see that the first-reflection component already demonstrates fluctuating AM
 and IF and presents a challenge for data analysis. Therefore, we shall use the
 synthesized $r(t)$ to help determining ConceFT parameters empirically. Secondly,
 the global-symmetry assumption happens to allow certain approximations 
 that lead a simpler expression for $r(t)$. It turns out that
 the level of temporal fluctuation in the AM and IF of $r(t)$ closely depends on 
 the ratio $\Delta x/\Lambda$. Details of mathematical derivation are given in Appendix B, C, and D, and related discussion will be given in Sec.~\ref{sec:new4}.
 }

\subsubsection{Choosing the SST and MT parameters}
An example of $R(\omega)$ was synthesized via Eq.~(\ref{eq:1}) with the following parameters: $l = 0.72$ cm, 
$\Lambda = l/5.5$ \citep{SheraBergevin-2012}, and $\Delta x = \Lambda/2$. 
The irregularity function $\epsilon(x)$ was generated by 
a zero-mean Gaussian random process with a constant variance $\text{E}[\epsilon(x)]^2 = \sigma_\epsilon^2$ and 
no spatial correlation; i.e.,  $\text{E}\{\epsilon(x) \epsilon(x')\} =0$ if $x \neq x'$. 
The integral with respect to $x'$ in Eq.~(\ref{eq:1}) was approximated by discretizing along the $x'$ direction from
$x'=0$ to $35$ mm with a step size of $5$ $\mu$m. The resulting $R(\omega)$ was calculated in the frequency domain from 0.2 to 16 kHz and the inverse
fast Fourier transform was applied to obtain the TEOAE signal $r(t)$ sampled at $f_s = 32$ kHz. Then, $\mathfrak{Re}\{r(t)\}$ was subject to further analyses. To test the robustness
of the signal-processing methods, Gaussian white noise $\xi(t) \sim \mathcal{N}(0,\sigma_\xi^2)$ was added to $\mathfrak{Re}\{r(t)\}$. 
Figure 1 shows $\mathfrak{Re}\{r(t)\}$ before and after contaminated by the noise. 
This signal is further analyzed as below. 

\begin{figure}[t]
  \centering
    \includegraphics[width=0.55\textwidth]{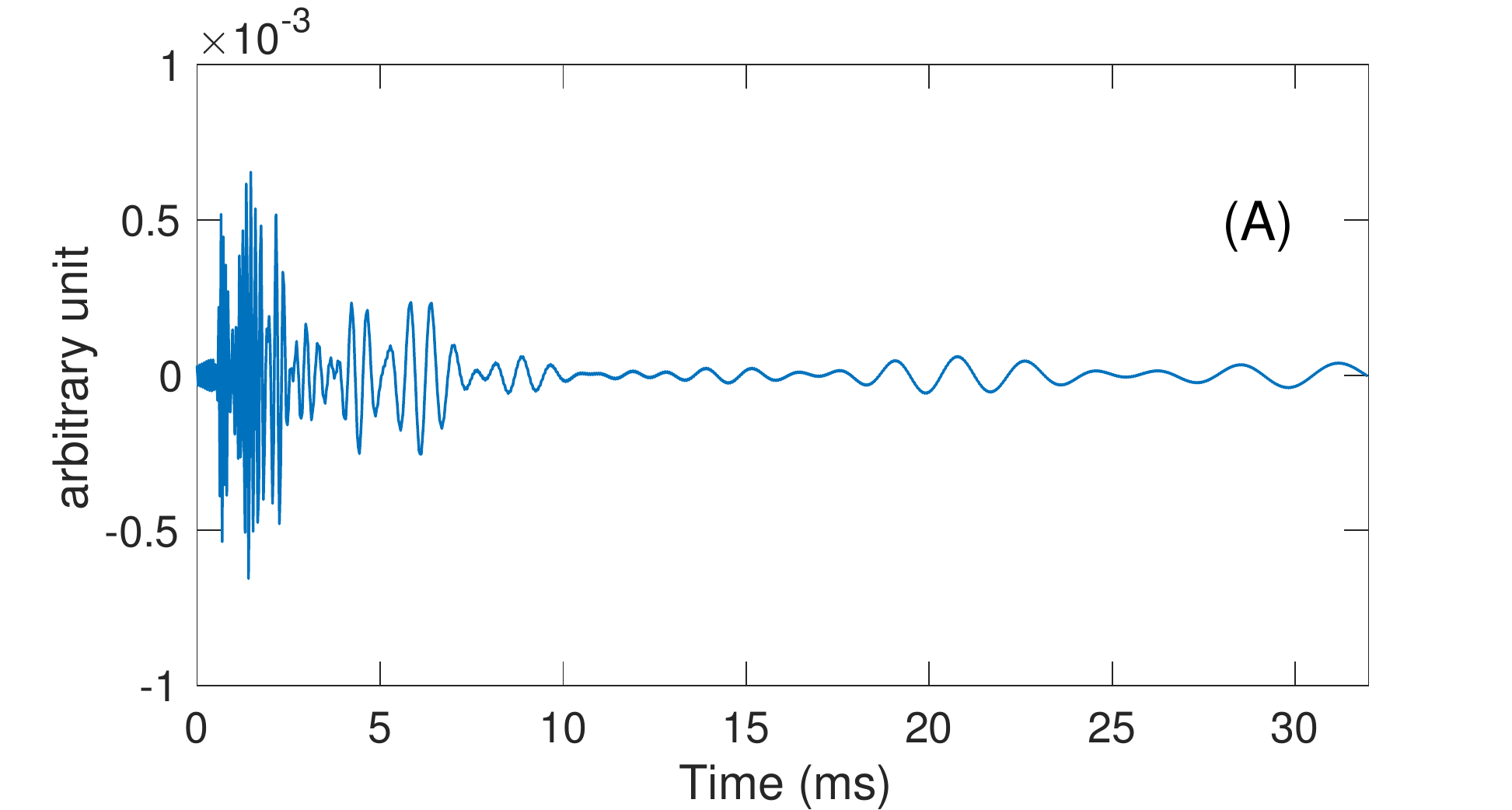}
    \includegraphics[width=0.55\textwidth]{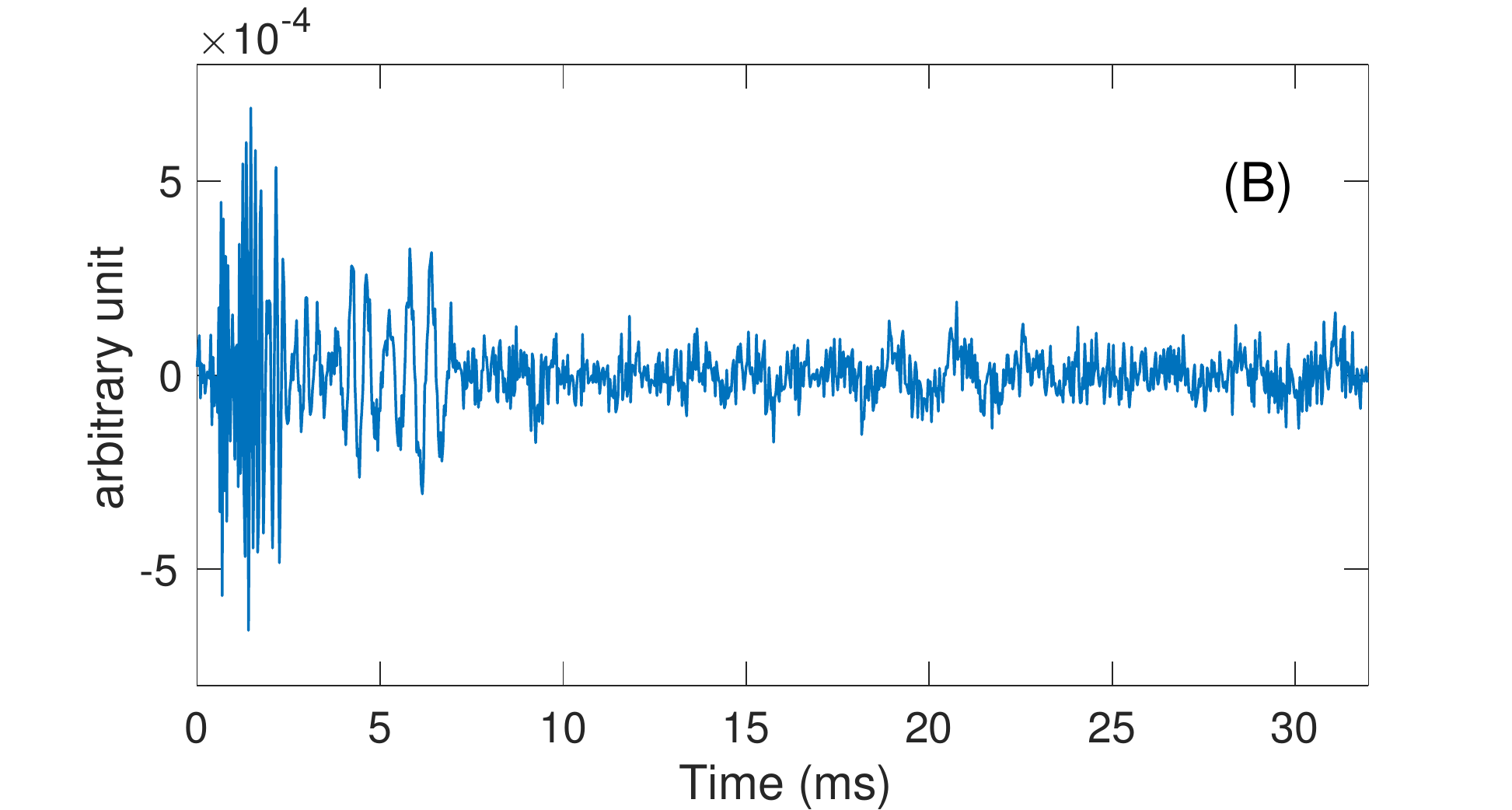}
    \caption{Signals generated by Eq.~(\ref{eq:1}) for further analysis. (A) the real part of the clean TEOAE signal $r(t)$. (B) Contaminated by
    Gaussian noise with $\sigma_\xi = 4.7\times 10^{-5}$. }
    \label{fig:1}
\end{figure}

\begin{figure}[htb]
  \centering
    \includegraphics[width=0.45\textwidth]{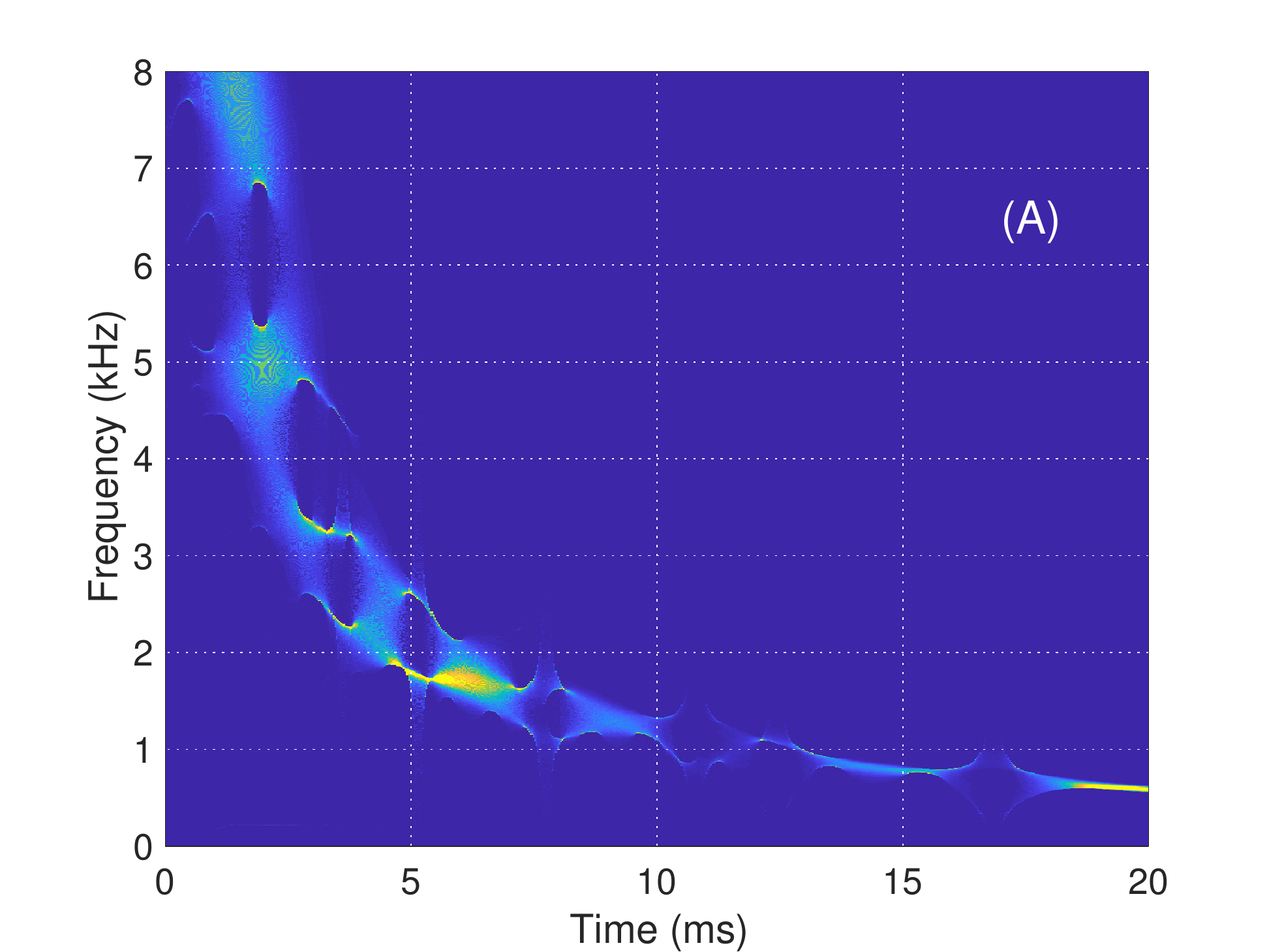}
    \includegraphics[width=0.45\textwidth]{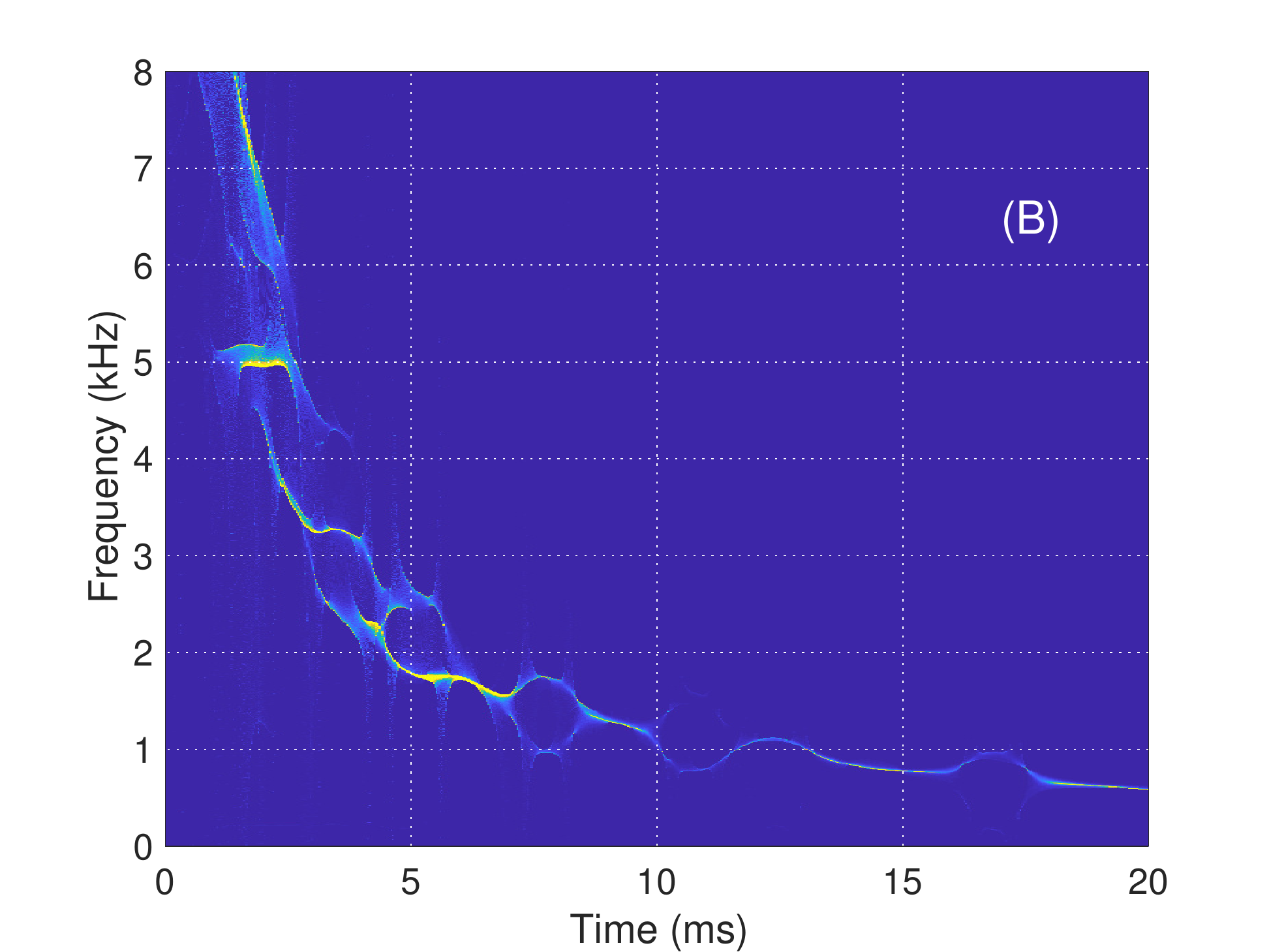}
    \includegraphics[width=0.45\textwidth]{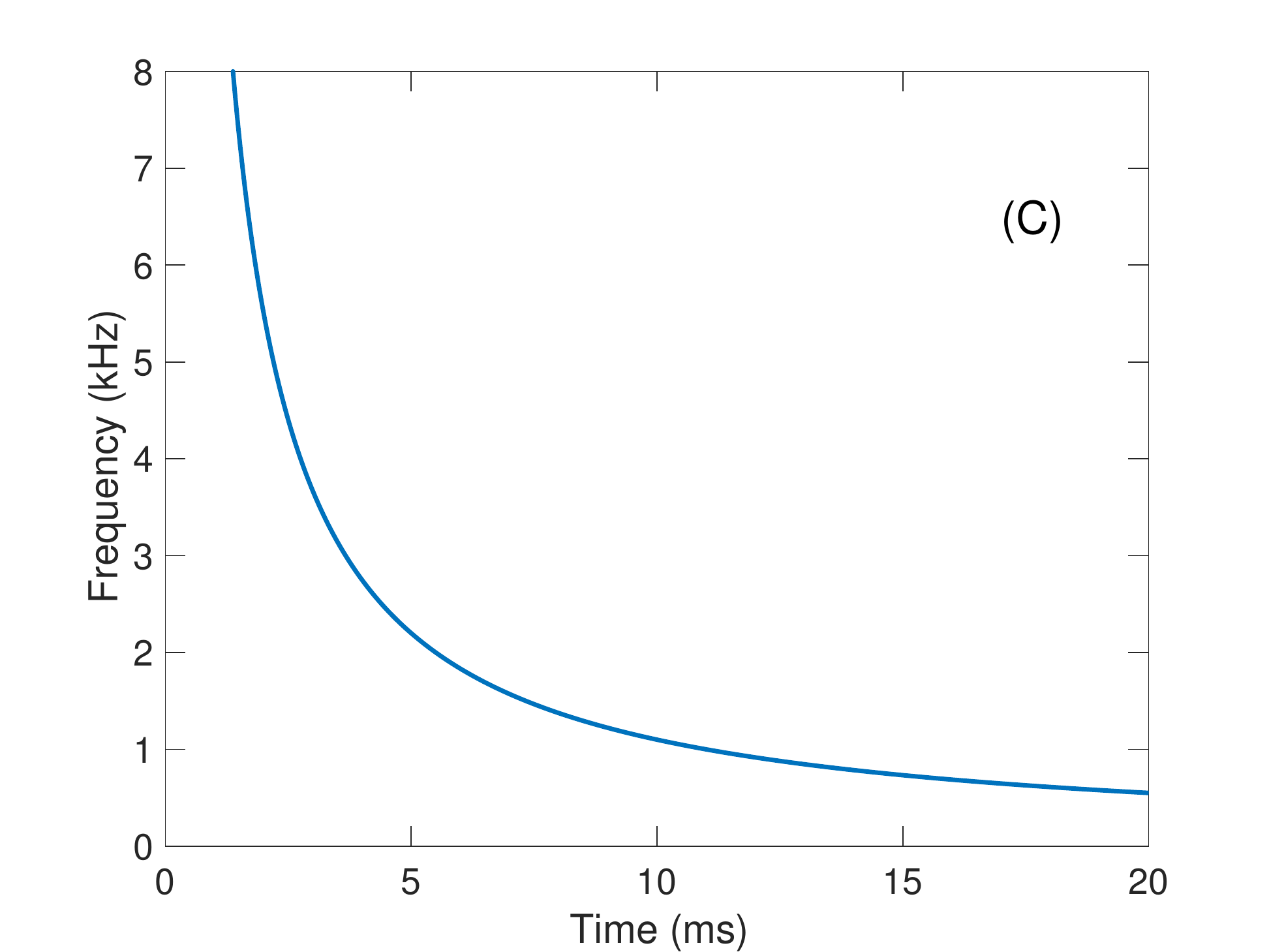}
    \includegraphics[width=0.45\textwidth]{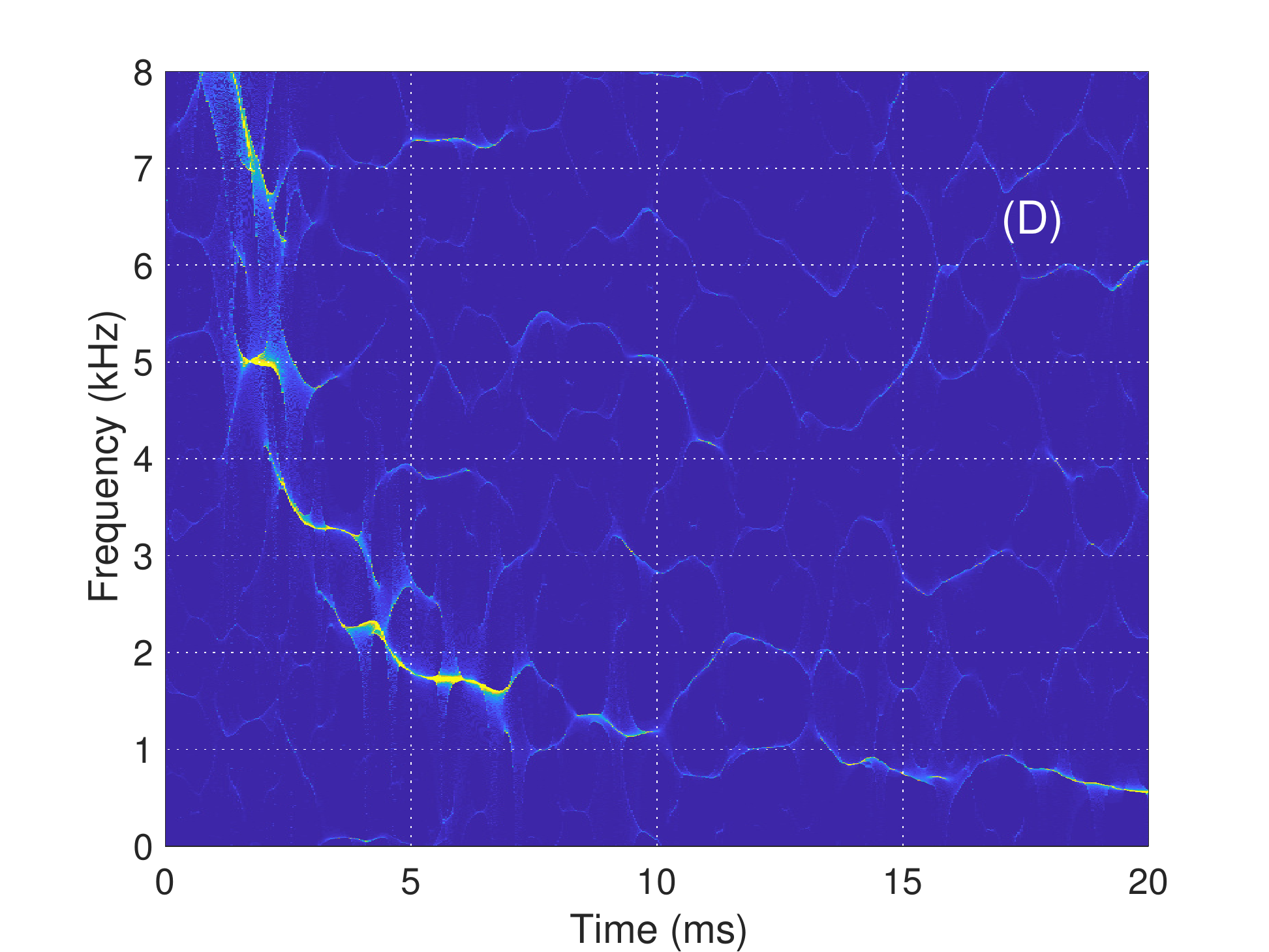}
    \includegraphics[width=0.45\textwidth]{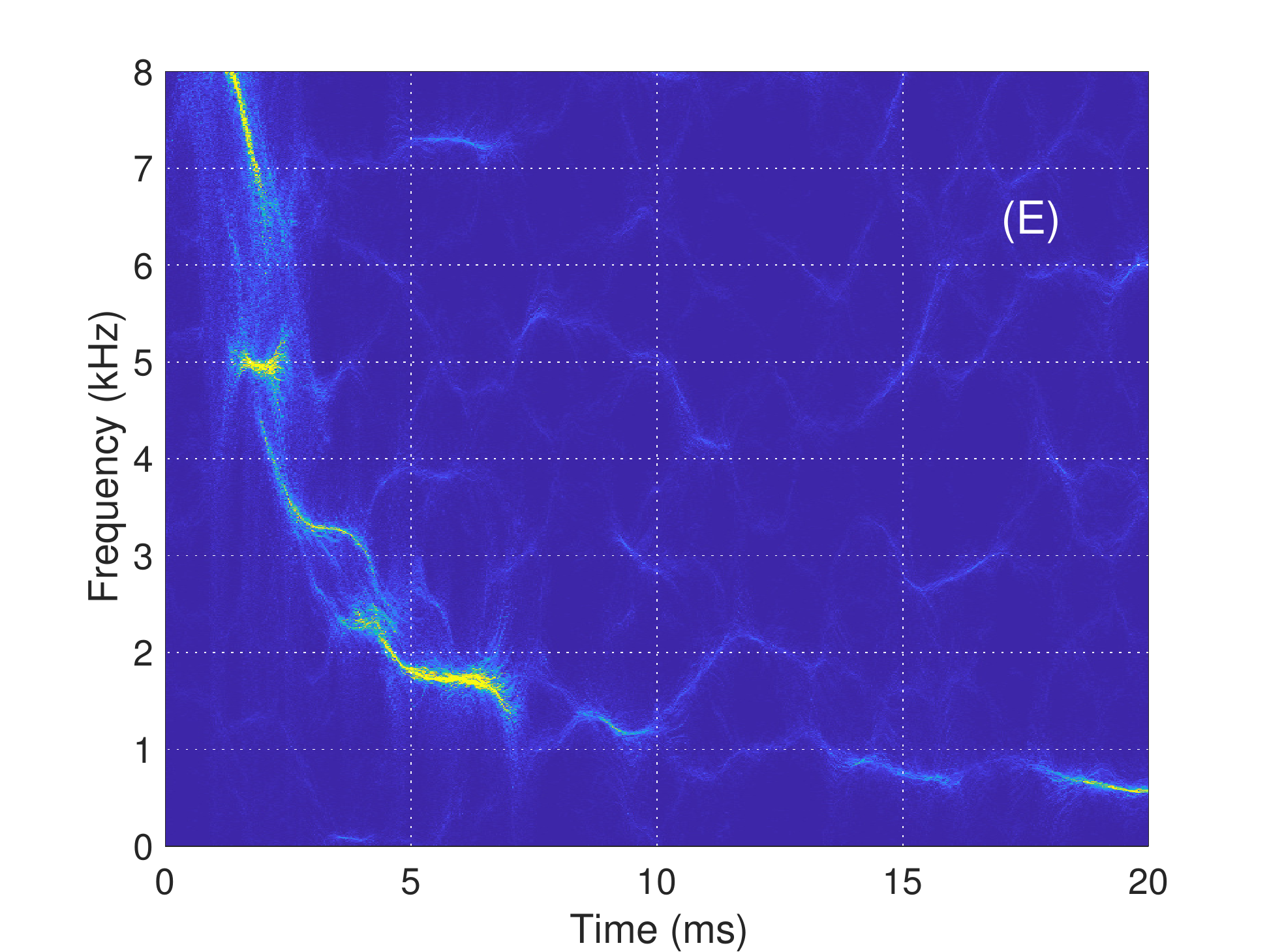}
    \includegraphics[width=0.45\textwidth]{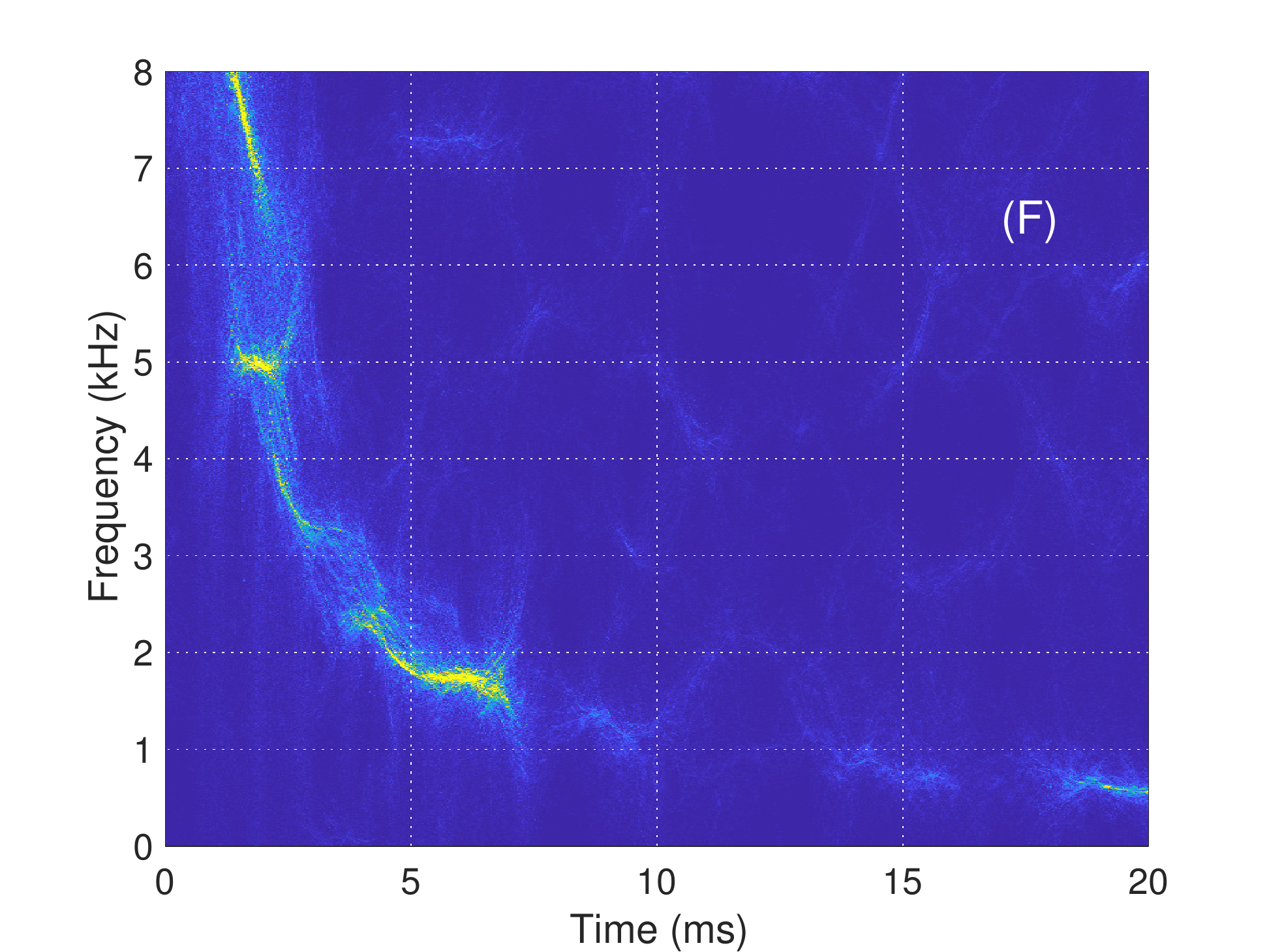}
    \caption{(Color online) Results of representing simulated TEOAE signals by SST and ConceFT. {\bf(A)} 1st-order SST, clean signal. 
    {\bf(B)} 2nd-order SST, clean signal. 
    {\bf(C)} The \emph{expected} instantaneous frequency (see Eq.~\ref{eq:IF} for 
    the definition) as a function of time.
    {\bf(D)} 2nd-order SST, noise contaminated. {\bf(E)} ConceFT with $J=2$ and $N=90$. {\bf(F)} ConceFT with $J=3$ and $N=90$.}
    \label{fig:2}
\end{figure}

An STFT spectrogram $V_f^{(h)}$ of signal $f(t) = \mathfrak{Re}\{r(t)\}$ was calculated using the Gaussian window
function $h(t)$ with $\sigma = (5 \text{ms})/12$. Then, both the 1st-order and the 2nd-order SST were tried
for comparison purposes, and the resulting T-F representations
$S_f^{(h)}(t,\nu)$ are shown in Fig.~\ref{fig:2}A and B. 
The \emph{expected} instantaneous frequency (EIF) function 
is plotted in Fig.~\ref{fig:2}C as a reference; 
The EIF function, denoted as $\bar{\nu}(t)$, is essentially defined 
as the inverse function of the expected group delay\footnote{In human data, the inverse relation was
find to hold for a limited period of time, from about $t=3$ to $8$ ms \cite{Keefe-2012}.} 
given by Eq.~(\ref{eq:Egd}); to be exact, we define that
$\text{E}\{\tau_g \big(2\pi\bar{\nu}(t)\big)\} = t$, and therefore,
\begin{equation}
\bar{\nu}(t) = \frac{1}{2\pi}\frac{4\pi}{t}\left(\frac{l}{\Lambda}\right) = \frac{2}{t}\left(\frac{l}{\Lambda}\right) = 11.0/t,
\label{eq:IF}
\end{equation}
where the unit of $t$ is sec and the unit of $\bar{\nu}$ is Hz.
 
Note that in Fig.~\ref{fig:2}A and B, both the 1st and the 2nd order SST produce
traces that generally follow Eq.~(\ref{eq:IF}) but with blurring and deviation. The background is clean in both cases because
the noise has not been added yet. Also note that, compared to the 1st-order SST, the trace in the 2nd-order SST 
appears to be more concentrated {when the IF changes fast}, {as was predicted by the established theorem \cite{Daubechies_Lu_Wu:2011,Oberlin2015}.} For this reason, the 2nd-order SST is chosen as the tool for analyzing the signals further.

Fig.~\ref{fig:2}D shows the 2nd-order SST of the contaminated signal $f(t) = \mathfrak{Re}\{r(t)\} + \xi(t)$ (See Fig.~\ref{fig:1}B).
By inspection and comparing with Fig.~\ref{fig:2}B, we can find several spurious traces in the upper-righthand side
of the ``main trace'' due to the additive noise (the main trace is vaguely defined as the set of all visible traces located 
close to $\nu= \bar{\nu}(t)$ on the T-F plane). Fig.~\ref{fig:2}E and F show the  result of further processing by ConceFT
using the first $J=2$ and $3$ Hermite basis functions, respectively. The number of averages in Eq.~(\ref{eq:C_y}) is $N=90$. 
Note that, before time $t = 10$ ms, the  
the main trace appears to be preserved in both Fig.~\ref{fig:2}E and F.
At time $t > 10$ ms, however, the main trace appears to be harder to identify, especially between $t = 10$ and $13$ ms. 
{In fact, the global scaling symmetry assumption in this model predicts that the TEOAE amplitude would decay fast;
and we attempt to provide a mathematical explanation for the reasons in Appendix \ref{sec:D}. Consequently, it becomes harder to see the tracing at later time as
the signal level eventually drops beneath the noise floor.}

Also note that the spurious traces due to the presence of noise appear to be suppressed better when using $J=3$ Hermite
functions (Fig.~2F) than $J=2$ (Fig.~2E), and this improved performance possibly comes with the price of lowering 
the visibility of the main trace when the SNR is below a certain limit, 
e.g., around $t=12$ and also $t=17$ ms, respectively.

\subsubsection{Results with a more realistic value of $\Delta x/\Lambda$}
\begin{figure}
\centering
\includegraphics[width=0.5\textwidth]{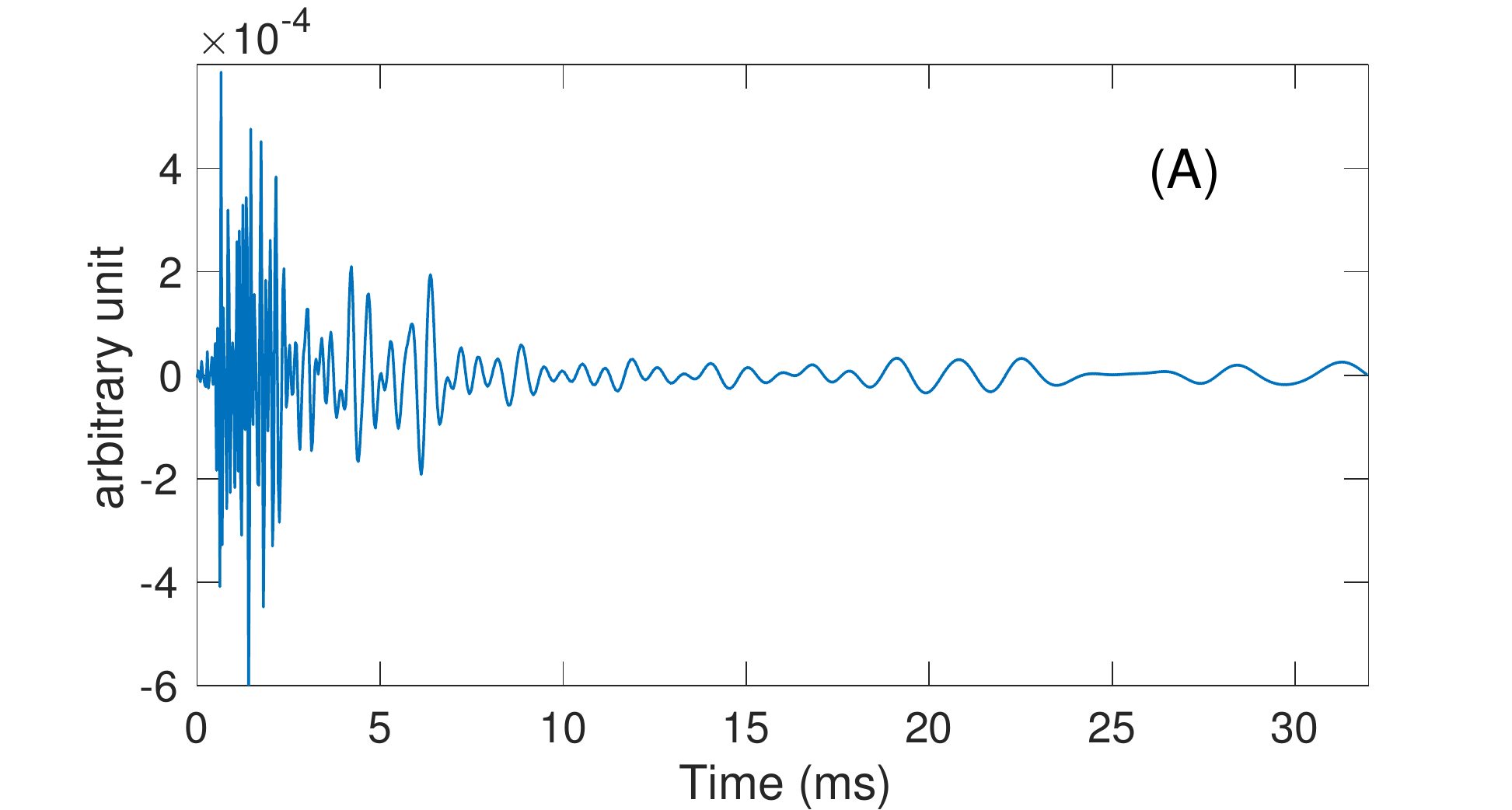}
\includegraphics[width=0.47\textwidth]{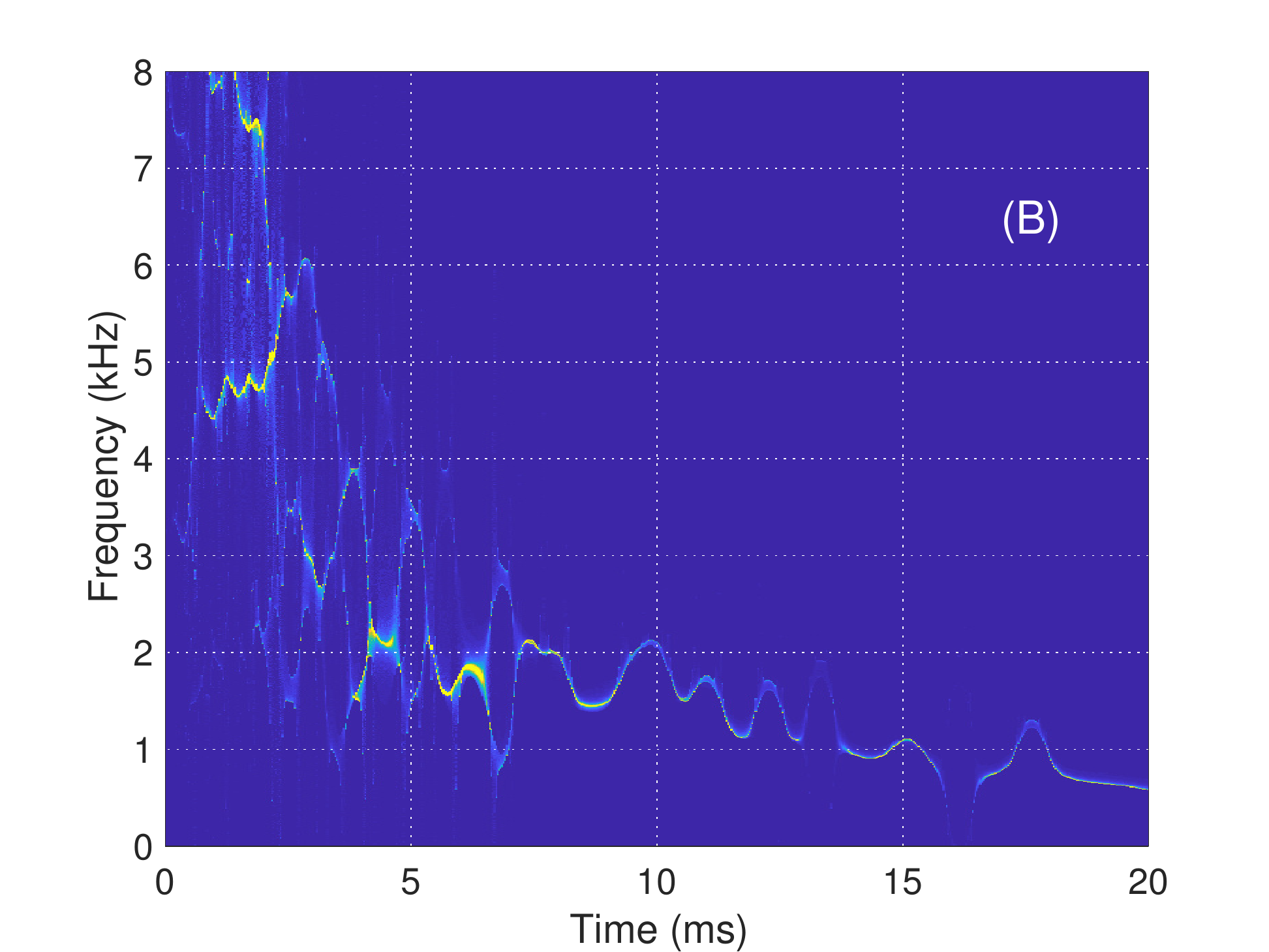}
\caption{(Color online). Results with a more realistic value of $\Delta x$. 
{\bf(A)} $r(t)$ with $\Delta x = l/(\sqrt{2\pi} 10) \approx 0.22 \Lambda$, as in \cite{SheraBergevin-2012} and \cite{BiswalMishra-2017}.  {\bf(B)} the corresponding 2nd-order SST, obtained with a Gaussian window 
$\sigma = 0.167$ ms.} 
\label{fig:5}
\end{figure}
Based on Eq.~(\ref{eq:1}), we will show in Appendix \ref{sec:B} and \ref{sec:C} that the OAE signal, $b(t)$, 
evoked by a narrowband stimulus $g(t)$ that is centered around frequency $\omega_b$ could be approximated as follows,  
\begin{equation}
	b(t) \approx C'' R(\omega_b) \cdot g\left(t-\frac{4\pi}{\Lambda}\frac{l}{\omega_b}\right),
\label{eq:14}	
\end{equation}
where $C'' = e^{i\frac{4\pi}{\Lambda}l}$ represents a constant phase shift that does not depend on $\omega_b$, and $R(\omega_b)$ depends on the irregularity function $\epsilon(x)$ is given by Eq.~(\ref{eq:1}). However, 
the goodness of approximation of Eq.~(\ref{eq:14}) 
relies on $\Delta x/\Lambda$ being sufficiently large, and it can be expected that the behaviors of 
TEOAE become more difficult to capture when $\Delta x/\Lambda$ is low. The goodness of approximation by
 Eq.~(\ref{eq:14}) and implications will be further discussed in Sec.~\ref{sec:new4}. 

Fig.~\ref{fig:5} shows the result of simulating
$r(t)$ with a more reasonable value of $\Delta x/\Lambda$  \citep{SheraBergevin-2012,BiswalMishra-2017}  inferred from experiments \citep{Rhode-1978}.  To simplify the discussion, no noise is added in this example. By inspection, the amplitude variation
 in Fig.~\ref{fig:5}A looks more irregular than Fig.~\ref{fig:1}A.
Moreover, the result of SST in panel (B) indicates that the IF also changes rapidly; in practice, this turns out to be difficult
to capture, and thus a shorter window was selected so as to obtain a clear plot.

\subsection{TEOAE from a cochlear model with electromotile outer hair cells}\label{sec:cochmod}

It has been shown in simulations that a computer cochlear model could generate TEOAE or SFOAE if 
random perturbation is introduced to some physical parameters along $x$ \citep{ChoiEA-2008, VerhulstEA-2012}. 
In this section, we adopt a model that captures certain bio-mechanical details of cochlear mechanics \cite{LiuNeely-2010} so that irregularities can
be placed in physically meaningful parameters. Then, by attaching a middle-ear model to the cochlear model { and terminating with an enclosed ear canal \cite{LiuNeely-2010}}, one can simulate
multiple reflections of the traveling waves so that the OAE signal becomes a superposition of multiple components, unlike
the single-component formulation described so far. We shall see that
the model also exhibits SSOAEs 
if the level of roughness is set sufficiently high. The simulated OAE data ``measured'' in the ear canal are then subject to
various ways of linear and nonlinear analysis so performance of different methods can be compared in Sec.~\ref{sec:comparison}.

The adopted model was based on an earlier transmission-line model of cochlear mechanics \cite{Neely-1985, Neely-1993} 
but the outer hair cell (OHC) ``modules'' were replaced by a piezoelectrical equivalent circuit \cite{MountainHubbard-1994, LiuNeely-2009} to account for OHC somatic motility. The mechanoelectrical transduction current of the OHCs were also made
to saturate so the entire system becomes nonlinear \cite{LiuNeely-2010}. However, in the past, the model parameters have been
intentionally designed to vary smoothly so it did not generate stimulus-frequency OAEs at a significant level \cite{LiuLiu-2016}. 
In the present research, we added roughness to the model by introducing randomness in the following way,
\begin{equation}
	m_b(x) = m_b^{(0)}(x) \big(1 + \epsilon(x)\big),
\end{equation}
where $m_b$ denotes the basilar-membrane (BM) mass density, $m_b^{(0)}(x)$ denotes
its mean values before introducing roughness, and $\epsilon(x)\sim \mathcal{N}(0,\sigma_\epsilon^2)$ as in 
Sec.~\ref{sec:cohref}. \citeauthor{ChoiEA-2008} showed that spectral filtering of $\epsilon(x)$ affects the level and spectral composition of coherent reflection-based OAEs. In the present work, once an instance of completely random $\epsilon(x)$ is generated, it is subject to moving-average smoothing so that the spatial correlation function $\kappa(x,x') = \text{E}\{\epsilon(x) \epsilon(x')\}$ is given as follows,
\[
\kappa(x,x') = \sigma_\epsilon^2 \cos^2\frac{\pi(x-x')}{2D},
\]
if $|x-x'| \leq D$, and $\kappa(x,x') = 0$ otherwise. 
\begin{figure}
\centering
\includegraphics[width=0.6\textwidth]{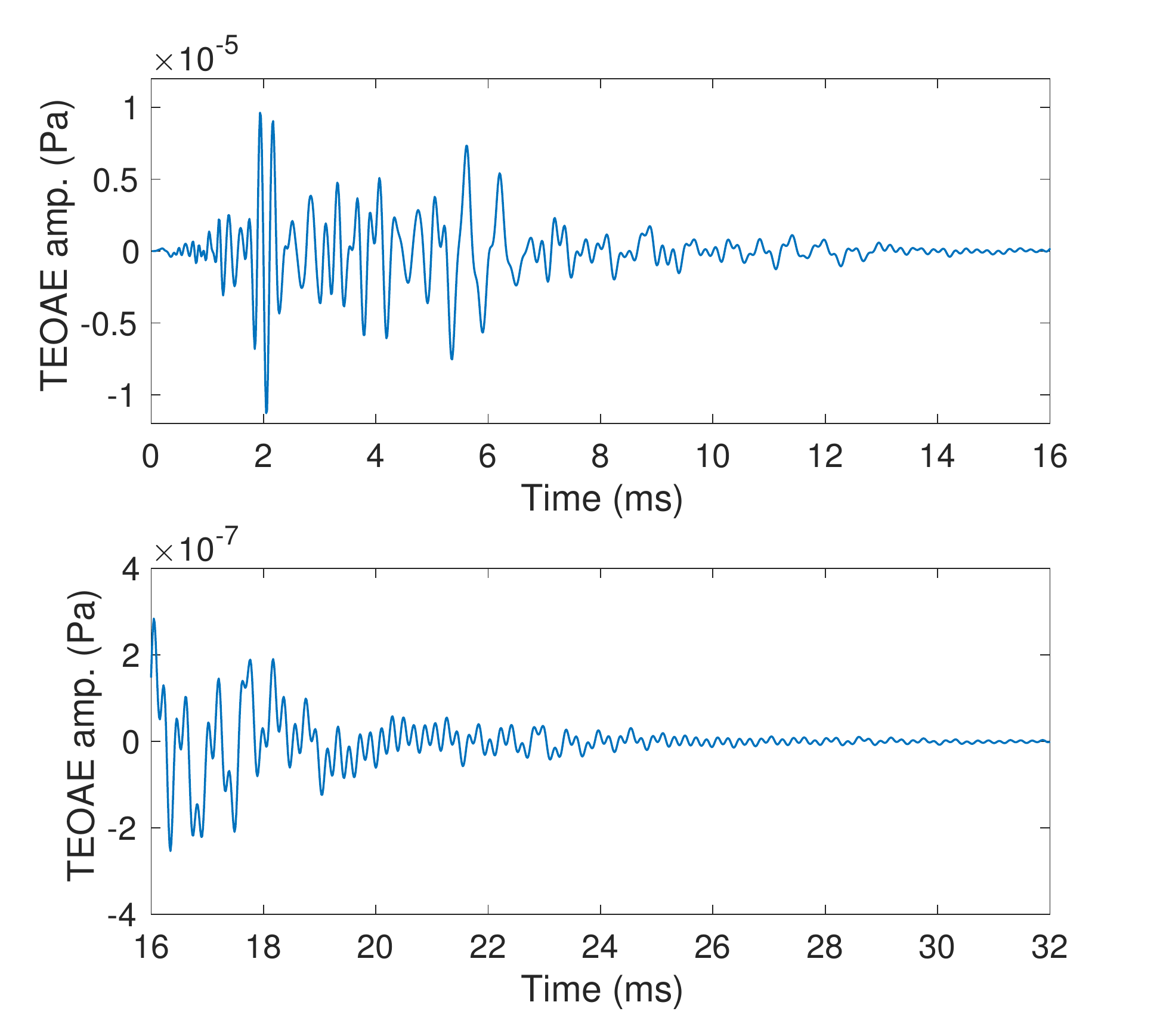}
\caption{An example of TEOAE obtained by introducing roughness to the \citeauthor{LiuNeely-2010} (2010) model. The stimulus is
a wide-band click with peak amplitude at $5.8$ mPa. The parameters for simulation are $\sigma_\epsilon = 0.03$, and $D = 0.3$ mm.}
\label{fig:TE-SSOAE}
\end{figure}
Figure \ref{fig:TE-SSOAE} shows an example of TEOAE obtained from this model, measured in the ear canal. The stimulus
is a wide-band click with a peak amplitude of 5.8 mPa. The stimulus actually generates ear-canal ringing for the first two
milliseconds or so. Therefore, we conveniently calculated the response with and without roughness in the cochlea, so that the ringing effect can be removed by taking the difference because it is mainly due to the linear responses of
the ear canal and the middle ear. What is shown in Fig.~\ref{fig:TE-SSOAE} is the difference signal, which can
be regarded as an accurate estimate of the true emissions since the original model is known to generate negligible TEOAEs without roughness \cite{LiuLiu-2016}.

The parameters for the roughness function are 
$\sigma_\epsilon = 0.03$, and $D = 0.3$ mm. Empirically, increasing $\sigma_\epsilon$ and decreasing $D$ tends
to increase the level of TEOAEs. In this particular example, apparently, there is a long-lasting high frequency component during $t = 16$ to $32$ ms, which can be made to disappear by using a lower value of $\sigma_\epsilon$ in simulation. Empirically, at $\sigma_\epsilon > 0.4$ the model
begins to produce self-sustained oscillation inside the cochlea and spontaneous OAE (SOAE) in the ear canal. We have not 
conducted a thorough search for the criteria for SOAE to happen in the model, as it is outside the scope of this paper.
The signal in Fig.~\ref{fig:TE-SSOAE} is used for comparing different T-F analysis methods, to be described next.

\subsection{Comparing different T-F representations}\label{sec:comparison}

\begin{figure}[htp] \centering{
\includegraphics[width=1.1\textwidth]{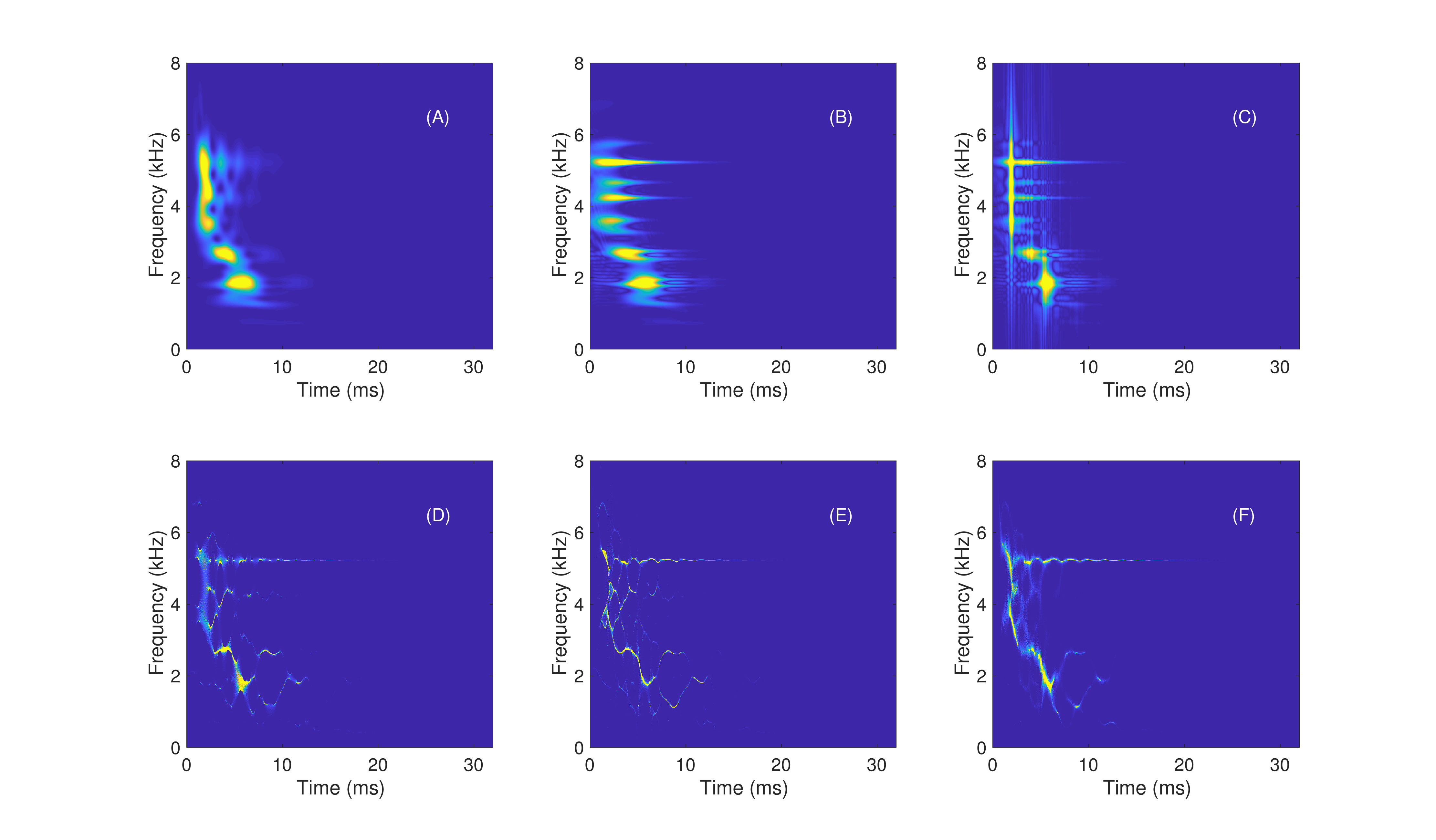}}
\caption{(Color online). T-F representations of a clean signal by different algorithms. The signal in Fig.~\ref{fig:TE-SSOAE}
is analyzed by (A) {scalogram}, (B) SPWV, (C) CWD, (D) {squared modulation $|s^{(h)}_{f}(t,\nu)|^2$ of the 1st-order SST, (E) squared modulation of the 2nd-order SST, and (F) squared modulation of ConceFT}, respectively.}
\label{fig:TFR-clean}
\end{figure}

\begin{figure}[htp] \centering{
\includegraphics[width=1.1\textwidth]{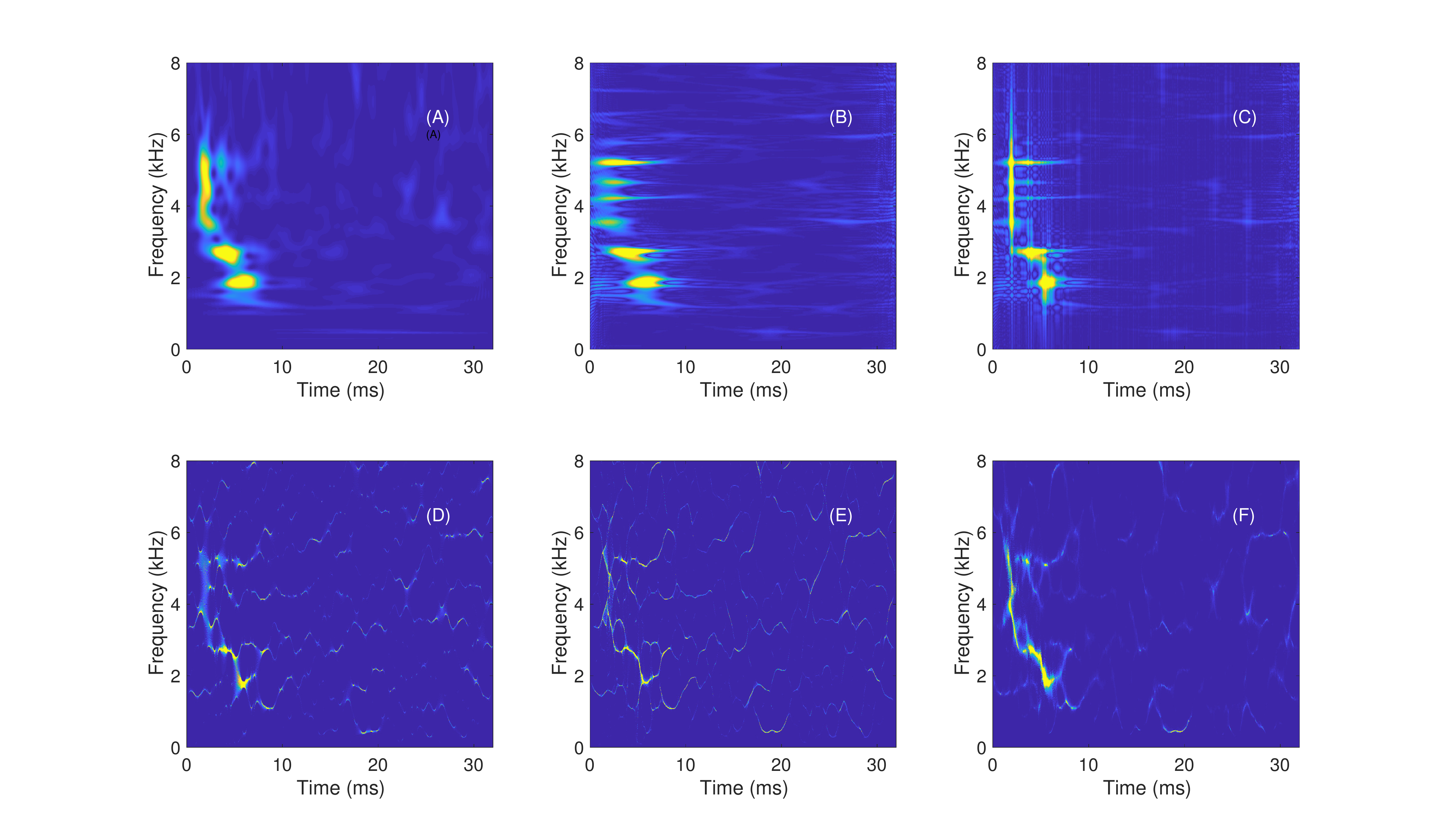}}
\caption{(Color online). T-F representations of the same signal as Fig.~\ref{fig:TFR-clean} but at SNR = 0 dB by (A) {scalogram}, (B) SPWV, (C) CWD, (D) {squared modulation $|s^{(h)}_{f}(t,\nu)|^2$ of the 1st-order SST, (E) squared modulation of the 2nd-order SST, and (F) squared modulation of ConceFT}, respectively.}
\label{fig:TFR-0dB}
\end{figure}

We compare ConceFT with commonly used T-F analyses, including STFT, CWT, CWD, and SPWV. For reproducibility, a publicly available toolbox called Time-Frequency Toolbox (TFTB) (\url{http://tftb.nongnu.org}) is used for the STFT, CWT, CWD, and SPWV. The code of ConceFT is available in the authors' website (\url{https://hautiengwu.wordpress.com/code/}). For the CWT, we use a suggested mother wavelet in previous papers \cite{Tognola1997, SistoEA2015}. For STFT, SST and ConceFT, we apply the same window length for a fair comparison. For STFT and SST, the Gaussian window is considered, and for ConceFT, the Gassian window and the first Hermite window is considered. For CWD and SPWV, the length of the time smoothing window is chosen to be the same as the window for STFT, and the length of the frequency smoothing window is chosen to be 2.5 times the length of the time smoothing window, as is suggested in the TFTB code. Since the scalogram (the squared modulation of CWT), CWD and SPWV are bilinear in nature, to have a fair comparison, the squared modulation of SST, { $|s^{(h)}_{f}(t,\nu)|^2$, and the squared modulation of ConceFT, $|C_Y(t,\nu)|^2$ 
(see Eq.~\ref{eq:C_y} for definition),} are displayed.

\subsubsection{Results of analyzing TEOAE and SSOAE generated by the \cite{LiuNeely-2010} model: visual comparison}
The results of analyzing the signal in Fig.~\ref{fig:TE-SSOAE} by different methods are shown in Fig.~\ref{fig:TFR-clean} for visual comparison. All the T-F representation are able to capture the main trace dropping from 6 kHz
to 2 kHz and below in the first 10 ms. The main trace supposedly represents the first-reflection component of TEOAE. Beside the main trace, a few other traces are noteworthy in SST and ConceFT; first, we see that a trace near 5.0 kHz can be easily captured in SST or ConceFT. The trace extends over 20 ms and is especially visible in ConceFT{, and it
certainly corresponds to the long tail in Fig.~\ref{fig:TE-SSOAE} which extends to time $>$ 25 ms. We refer
to this long-lasting component as the SSOAE here.}  In contrast, in the {scalogram}, SPWV, and CWD, the component is not as easy to identify. Note that the 5-kHz component in the {scalogram} looks like ``wideband''. This is because in the frequency domain the mother wavelet is wide in the high frequency region {due to the dilation nature of the CWT.}

Secondly, to the right of the main trace in SST and ConceFT, we arguably see a second trace at doubling the time.
For instance, near $t=8.4$ ms a component at $2.7$ kHz re-occurs (after its first occurrence near $t=4.2$ ms), 
and arguably that component drops to $1.8-1.9$ kHz at around $t = 11.4$ ms. Based on hindsight, since the 
\citeauthor{LiuNeely-2010} model consists of the middle-ear part and the cochlear part, it should not be
surprising to see multiple reflections due to the impedance mismatch at the stapes. Hence, this second trace at
doubling the latency likely represents the second-reflection component of TEOAE. Note that the component 
is also visible in {the scalogram}, as it has been reported when analyzing synthetic data generated
by a model with internal reflection \cite[their Fig.~10]{SheraBergevin-2012}; previously, the linearity of CWT conveniently allowed separation of reflection components by masking out part of the T-F representation and 
applying the inverse transform. The component, however, seems not so visible in bilinear transforms (SPWV and CWD) for this particular example.

To further examine the performance of ConceFT at low SNR region, we add a Gaussian white noise to the signal, with the SNR = 0 dB calculated over the entire time (32 ms). The result is shown in Fig.~\ref{fig:TFR-0dB}. As can be visualized, even when the SNR is so low, the SSOAE component can still be identified with ConceFT. 
Although the SSOAE component now appears to be blurred in ConceFT, it is still more ``concentrated'' than
in the CWT. The main trace remains robust against the noise in all 6 representations, while the trace appears
brighter in ConceFT than in SST. 
Comparing (D) and (E) to (F), we also see that the noise seems to be more successfully ignored by ConceFT than
by 1st-order and 2nd-order SST, which demonstrates the effectiveness of multi-tapering in providing robustness against 
additive noise. 

\begin{table}[t]
\caption{Recovery of the iTFR by different T-F analysis methods. Performance is evaluated by the optimal transport distance (a smaller distance means a better recovery). {The mean and standard deviation of 30 realizations of noise is reported. SPWV: smooth pseudo Wigner Ville distribution; CWD: Choi-Williams distribution; 1-st SST: the squared modulation of the first order SST; SST: the squared modulation of the second order SST; ConceFT: the squared modulation of concentration of Frequency and Time.}}
\begin{center}
\begin{tabular}{r || c | c | c | c |c |c}
\hline
SNR &{\bf {scalogram}}  	&{\bf SPWV} 		&{\bf CWD} & {\bf 1st-SST} 	&{\bf SST} &{\bf ConceFT} \\
\hline
100 dB & 2.27 (0.00) &2.03 (0.00) &2.93 (0.00) &0.69 (0.00) &0.83 (0.00) &1.20 (0.07) \\
10 dB & 3.29 (0.11) &3.34 (0.12) &3.55 (0.08) &2.65 (0.13) &2.75 (0.12) &2.61 (0.14) \\
5 dB & 3.53 (0.11) &3.61 (0.14) &3.73 (0.09) &3.02 (0.14) &3.11 (0.13) &3.00 (0.14) \\
2 dB & 3.65 (0.11) &3.73 (0.12) &3.84 (0.08) &3.23 (0.13) &3.30 (0.12) &3.21 (0.14) \\
0 dB & 3.72 (0.11) &3.81 (0.11) &3.90 (0.09) &3.36 (0.13) &3.42 (0.13) &3.35 (0.14) \\
\hline
\end{tabular}
\end{center}
\label{tab:OTD}
\end{table}%

\subsubsection{Results of analyzing a three-component, OAE-like synthetic signal: Comparison by optimal transport distance}
To facilitate quantitative comparison between the performance of different T-F analysis methods, we synthesized an OAE-like signal { with the following specification}. 
Let $F_{a,b,c}(t)=a+\frac{S_c\{W(s)\}}{b\max_{0\leq s\leq L}(S_c\{W(s)\})}$ denote a time function with length $L$, mean $a>0$ and $b>0$ and perturbed by
the standard Brownian motion $W(s)$ with $W(0)=0$ \cite{Yakov:book}, and $S_c$ is the locally weighted smoothing operator of kernel bandwidth $c>0$.
A signal consisting of three oscillatory components was produced, each with a time-varying amplitude and frequency. The first component is a chirp-like signal with frequency dropping {from 8000} to 2000 Hz following the $1/t$ rule, {which lasts from 1 ms to 20 ms. This component} simulates the TEOAE component. The phase {$\phi_1(t)$ is a realization of} $1000(2\pi)\times (\frac{120}{19}\log(t/\text{1ms})+\frac{13}{19}(t-1)+F_{1,6,0.3})$.
The second component oscillates around frequency 5000 Hz, {which lasts from 2ms to 25 ms. This component} may simulate an SSOAE-like component. The phase {$\phi_2(t)$ is a realization of} $1000(2\pi)(5t+0.1\cos(\pi t)+F_{0,5,0.4})$.
The third component oscillates around 3141 Hz, which lasts from 3 ms to 10 ms. It may simulate  another
SSOAE-like component. The phase is $\phi_3(t)= 3141(2\pi) t$.
The amplitude of the three components, $A_1(t)$, $A_2(t)$ and $A_3(t)$, are realizations of $F_{1,2,0.2}$, $F_{1/2,4,0.1}$, and $F_{1/3,6,0.1}$ respectively.  Note that a realization of $F_{a,b,c}$ is a smooth function and due to its random nature it is not { easy} to express it by any well known function,
which makes this evaluation somewhat realistic.

The signal is sampled at 32000 Hz, and the SNR ranges from 100 to 0 dB. We then apply ConceFT, {SST,} CWT, CWD, and SPWV to the simulated signal. To evaluate the performance of different T-F analyses on this signal, we define the ``ideal T-F representation (iTFR)'' in the following way --- suppose the signal is $f(t)=\sum_{l=1}^3A_l(t)e^{i\phi_l(t)}$. The iTFR is defined as follows,
{
\begin{equation}
R(t,\omega)= \sum_{l=1}^3 A_l(t)\delta(\omega-\phi_l'(t)),
\end{equation} 
where $\delta(\omega)$ denotes the Dirac delta distribution. 
Ideally, we would like to recover iTFR as much as possible. 

To assess the performance of different algorithms in recovering the iTFR, {we follow the suggestion in \cite{Daubechies-EA-2016}} and calculate the optimal transport distance (OTD) between the iTFR and each T-F representation, respectively. 
{The OTD is sometimes called the earth mover distance, and is associated with the Monge's optimal transport problem \cite{Villani:book}, which provides a way to measure similarity between probability distributions. 
}
Numerically, the OTD can be calculated in the following way \cite{Villani:book}:
for two probability measures $\mu$ and $\nu$ defined on $\mathbb{R}$, let $f_\mu(x)=\int_{-\infty}^x\, \ud \mu $  { denote the cumulative distribution function of $\mu$} (and analogously for $f_{\nu}$). Then, the OTD is defined as
\begin{equation}
d_{\mbox{\footnotesize{OT}}}(\mu,\nu)= \int_{\mathbb{R}} |f_\mu(x)-f_\nu(x)|\, \ud x\,.
\label{eq:OTD}
\end{equation}

\begin{figure}
\centering
\includegraphics[width=0.7\textwidth]{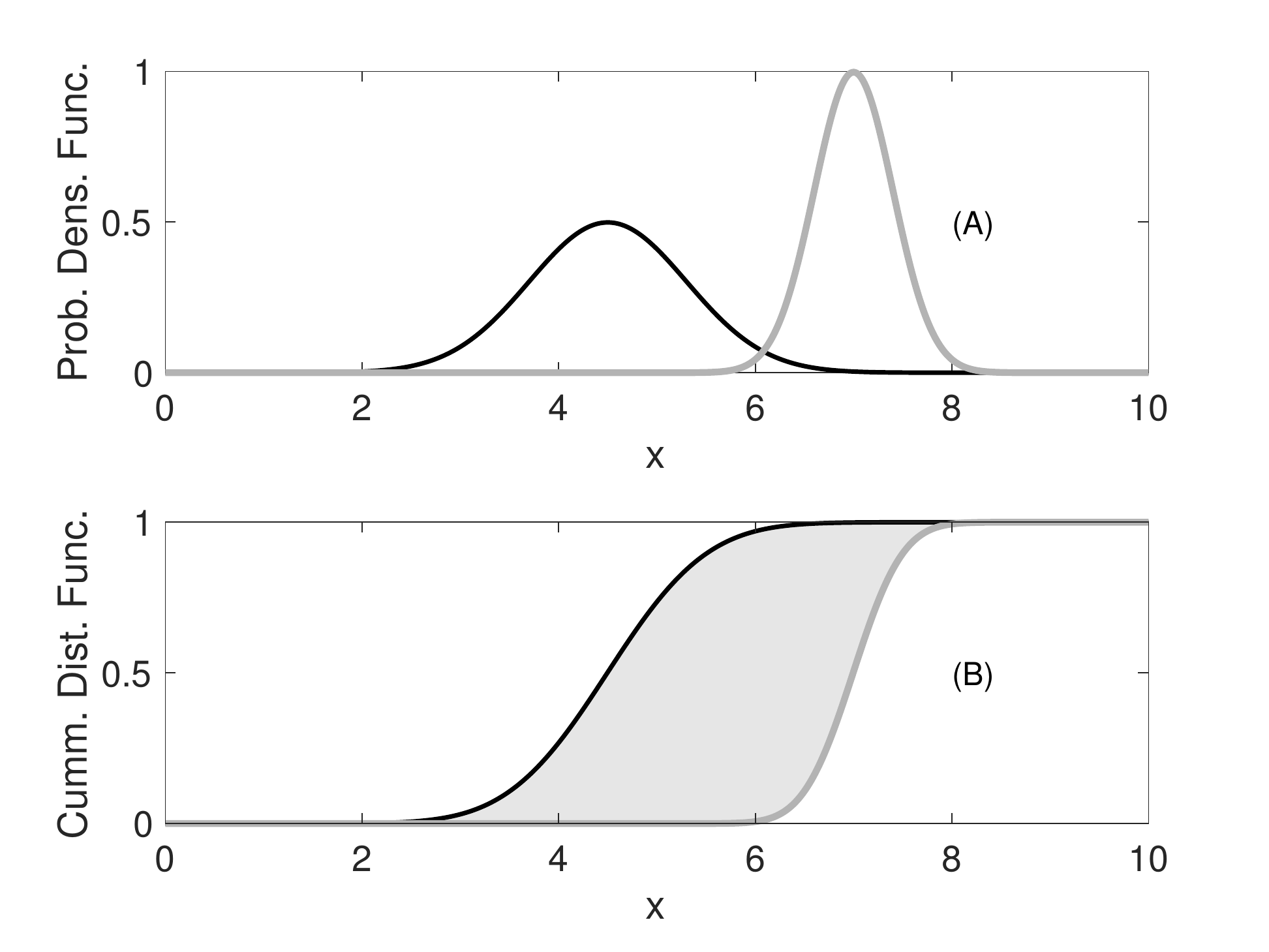}
\caption{Illustration of the optimal transport distance (OTD). In (A), the black and the grey lines 
show two probability density functions (PDFs), respectively, and their corresponding cumulative distribution functions (CDFs) are shown in (B). The shaded area shows the OTD
 between these two probability measures.}
\label{fig:OTD}
\end{figure}

{
In the plain language, the OTD is the minimal amount of ``effort'' (in the unit of mass times distance) that
is required to transfer an amount of mass from several locations to other locations. 
Fig.~\ref{fig:OTD} illustrates this idea, and the OTD equals to the area bounded between two cumulative distribution functions, as it is
defined in Eq.~(\ref{eq:OTD}). Note that the OTD not only captures the distance (in the $x$-direction) between the peaks of distributions, but is also affected by the width of the distributions.

In the context of comparing
different T-F representations against the ideal one (which has an infinitely narrow width), the OTD thus allows us not only to measure the degree of concentration of each T-F representation, but also its correctness in estimating the true IF. To evaluate the goodness of T-F representations for TE or SFOAEs, previous work has used the mean or mean-square distance between the maximum location and the ground-truth location as the performance metric \cite{SheraBergevin-2012, BiswalMishra-2017}. This kind of performance metric naturally requires a pre-filtering step in the T-F plane to ensure that
only the first-reflection component remains in the T-F plane. Note that OTD can be viewed as a generalization of such performance metric which uses all density information for the purpose while not requiring pre-filtering. 
Thus, the OTD is chosen here as a way to evaluate how accurately the concentration of the iTFR is captured by different algorithms.}

Specifically, for each fixed time $t$, the T-F representation obtained via each algorithm is treated as a probability measure on frequency $\omega$. 
Note that in general at each time $t$, the T-F representation does not have integral $1$. Thus, the T-F representation is normalized first. 
Then, the OTD between the obtained T-F representation and the iTFR at time $t$ is calculated, and its average over all time is reported as a measurement of accuracy of analysis.
The results of analyzing the afore-mentioned signal by different methods, {each with 30 realizations of noise,} are reported in Table \ref{tab:OTD} in terms of the OTD.

\section{Discussion}\label{sec:new4}
In this section, we would like to suggest a view based on Eq.~(\ref{eq:1}) and the approximation given in Eq.~(\ref{eq:14}) to argue that TEOAE is intrinsically hard to analyze even when just considering the 
first-reflection component $r(t)$. We wish to propose that the expansion by IMT functions is thus an
appropriate way to model TEOAE signals when the situation is further complicated by internal reflection components
and SSOAEs. The performance of SST and  ConceFT has been reported in comparison with several 
existing methods in the field, so limitation of the present methods can be discussed, and a few future research directions will be pointed out.

\subsection{Why is TEOAE difficult to analyze? An insight from Eq.~(\ref{eq:14})}

The expression in Eq.~(\ref{eq:14}) predicts that a tone burst centered at a particular frequency $\omega_b$ is anticipated to
evoke an OAE component that returns around the time $t = \text{E}\{\tau_g(\omega_b)\} \propto \omega_b^{-1}$; 
further, the amplitude of that component would be scaled by a complex-valued gain 
$R(\omega_b)$. This result is rather simple to interpret, and since any transient and broad-band stimulus
 can be regarded as a superposition of narrow sub-band signals, Eq.~(\ref{eq:14}) essentially
 predicts that the first-reflection component of TEOAE will behave like amplitude modulated \emph{chirps}.  

However, the approximation relies on one crucial assumptions --- the 
wavenumber-domain representation $\tilde{\rho}(k)$ of the excitation pattern $\rho(x)$ must be sufficiently concentrated
around its peak $k = 4\pi/\Lambda$ so that Eq.~(\ref{eq:TB}) can be simplified. It is rather straightforward to 
show that the 2nd-moment of the function $\tilde{\rho}(k)$ is equal to $1/(\Delta x)^2$. Thus, similar to
how the Q-factor is defined in the frequency domain as the center frequency divided by the bandwidth, 
a $k$-domain factor can be defined for $\tilde{\rho}(k)$ and its value is $Q = 4\pi(\Delta x)/\Lambda$. A numerical investigation is given next to 
illustrate how the goodness of approximation in Eq.~(\ref{eq:14}) depends on $Q$.

\begin{figure}
\centering
\includegraphics[width=0.7\textwidth]{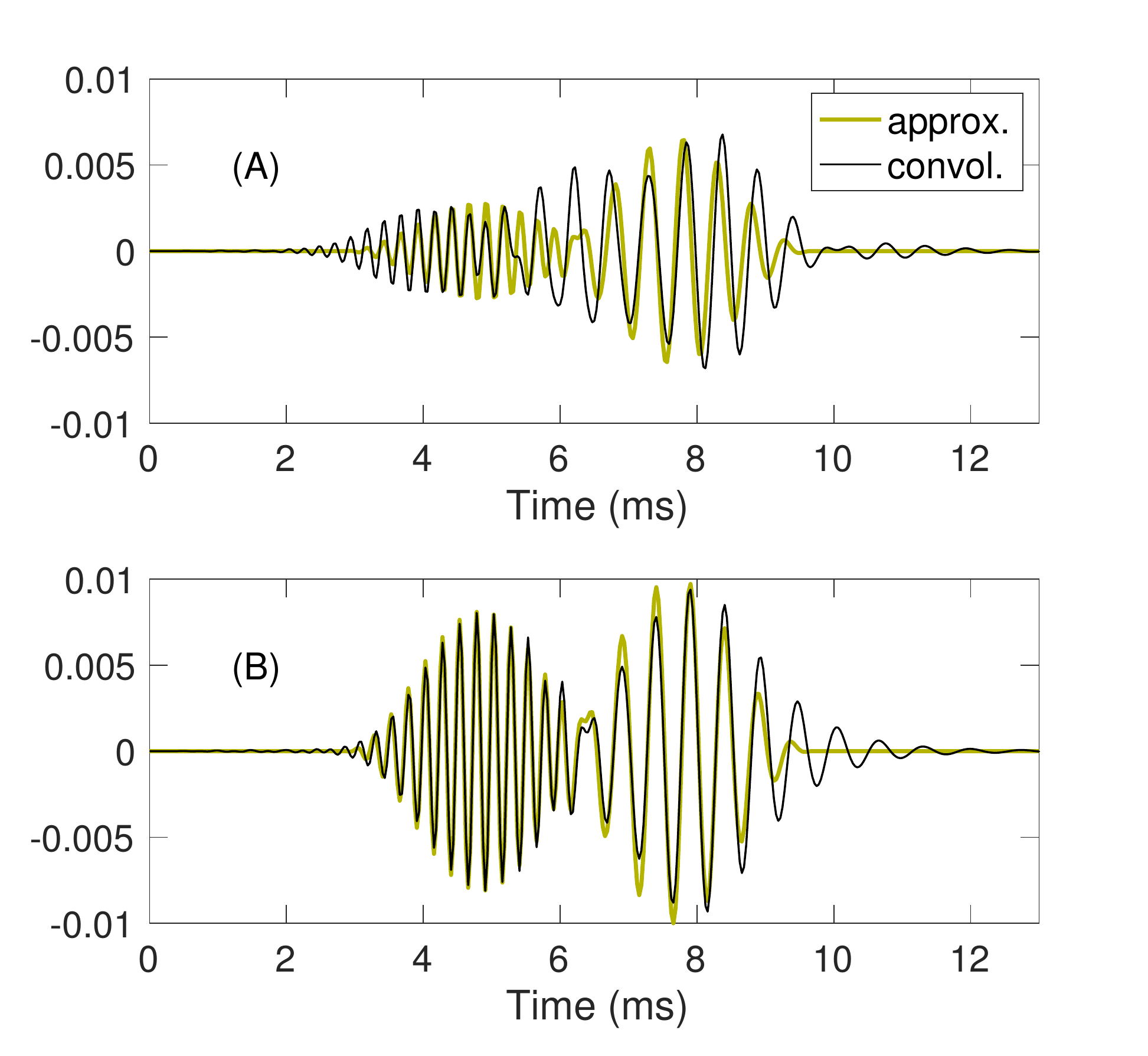}
\caption{(Color online) Accuracy of approximation 
by Eq.~(\ref{eq:14}) depends on the factor $Q = 4\pi(\Delta x/\Lambda)$. {\bf(A)} $\Delta x/\Lambda = 0.5$, the
default value in this paper. {\bf(B)} Goodness of fitting is better when $\Delta x/\Lambda = 2.0$.}
\label{fig:4}
\end{figure}

Here, a ``two-tone-burst'' stimulus $g(t)$ is prepared by applying a 4-ms Hann window to a mixture of two tones with
$\nu_1 = 4.0$ kHz and $\nu_2 = 2.0$ kHz; that is, $g(t) = h(t)\big(g_1(t) + g_2(t)\big)$ 
where $g_{1}(t) = \exp(i2\pi\nu_{1} t)$, $g_{2}(t) = \exp(i2\pi\nu_{2} t)$, and $h(t)$
denotes the Hann window with a support from $t=0$ to $4$ ms. Then, the precise first-reflection TEOAE component is calculated by $b(t) = g(t) * r(t)$, where $*$ denotes convolution in time. Additionally, an approximation $\hat{b}(t)$ based on Eq.~(\ref{eq:14}) is calculated as follows,
\begin{equation}
\hat{b}(t) = C''\sum_{j=1,2} R(2\pi\nu_j) \cdot g_j\left(t-\frac{2l}{\Lambda \nu_j}\right)
\end{equation}

Figure \ref{fig:4} compares $\hat{b}(t)$ and $b(t)$. Panel (A) is obtained by setting $\Delta x/\Lambda$ as $0.5$, a rather high value compared to $0.22$ suggested in the literature \cite[e.g.,][]{SheraBergevin-2012}, and panel (B) shows the result when $\Delta x/\Lambda = 2.0$, which is unreasonably high. The thin line (labeled as `convol.') shows $b(t)$ and the thick, lighter line (labeled as `approx.') shows $\hat{b}(t)$. In these particular examples the ratio $l/\Lambda$ is set to $5.7$ so we have $C'' = e^{i(0.8\pi)}$. 
By inspection, the approximation is better in (B) than in (A), and this agrees with the previous argument that
a higher $Q$ factor in the $k$-domain should result in a better approximation.

In Fig.~\ref{fig:4}(A), note that the peak of the 4-kHz packet seems to occur at an earlier time (near 4.0 ms) than predicted by
Eq.~(\ref{eq:14}) at $4.85$ ms, or $2l/(\Lambda \nu_1) = 2.85$ ms after the input packet peaks at $t=2.0$ ms. This can be regarded as an example of deviation of the group delay from 
its expected value, even though we have introduced a lot of crude simplifications to merely consider a single and clean reflection component from a 
globally scaling-symmetric and linear model. With a more realistic setting of $\Delta x/\Lambda$
at a lower value, one should expect the deviation to be larger. Thus, to model the signal appropriately,
 the expansion by IMT functions described in Sec.~{\ref{sec:new2}} perhaps provides better flexibility in characterizing what frequency components, single or multiple, are present at every moment. 

If we view the TEOAE signals as a sum of IMT functions, SST and ConceFT then could be
adopted to estimate the AM and IF of individual components.  Methods proposed in this paper might
also be helpful in extracting a robust T-F representation of real TEOAE data from individual ears with normal or impaired conditions. In particular, there has been increasing evidence that TEOAEs have significant short-latency components \cite{Goodman2009, JedrzejchakEA-2018} coming from locations that are basal to the characteristic places \cite{SistoEA2015}. 
Analyzing human data by SST and ConceFT is warranted as a future research topic.

\subsection{More about the present analysis method}

To compare the T-F representations in terms of their capability in preserving the ``ground-truth'' distribution,
Table {\ref{tab:OTD}} shows that the following results persist across different SNR levels: ConceFT is better than 
1st-order SST, followed by 2nd-order SST, CWT, and then the two bilinear methods SPWV and CWD. Results for the clean signal (SNR = 100 dB) is an exception, with SST outperforming ConceFT. {When the signal is clear, it is not surprising that ConceFT performs worse. This is because two windows are used, and the second Hermit window has a wider support in the T-F plane, and this slightly blurs the T-F representation of the final result. When noise is large, due to the MT effect, ConceFT performs better than SST. }

In \cite{BiswalMishra-2017}, CWT and SST were also compared among several other T-F representation methods (STFT, EMD, and S-transform). They found that the performance of CWT was better than SST at an SNR great than 15 dB in terms of group delay estimation error over a range of frequencies (0.4 to 8 kHz). The methodology of their work and the present work differ in several ways. First, the SNR range is different. Secondly, the test signal is different; they used a coherent reflection model with different realizations of the irregularity function to generate different instances of OAE signals with internal reflections, but here we choose to use signals with known ground truth directly. It turns out that their test signals had multiple-reflection components while the test signal here contains SSOAE-like components. Third, thus the performance measure is different; the group delay after Loess smoothing \cite{SheraBergevin-2012} was adopted as the true answer by \citeauthor{BiswalMishra-2017}, 
and T-F analysis methods were evaluated based on how closely
the smoothed group delay can be estimated despite of noise and internal reflections. Finally but not the least important, in their work, the SST was a reassignment of CWT, but in the present research the SST is based on STFT. {The difference between CWT-based SST and STFT-based SST deserves a discussion. While the dilation nature of CWT renders it suitable to better visualize the chirp like frequency latency structure of the OAEs, the phase information in the high frequency region is more mixed up compared with that of STFT {when there are multiple high frequency oscillatory components}. This is because to detect the high frequency component, the wavelet needs to be scaled down, and it is equivalent to broadening the frequency band. Due to the mixed up phase information in CWT, the reassignment result in the frequency axis is worse.\footnote{{When there are multiple high frequency components, the broadening frequency bands due to its dilation nature will cause mixed-up of different components, and hence the mixed-up phase information. This fact leads to the common ``dyadic separation'' assumption when we analyze multiple oscillatory components by CWT \cite{Daubechies_Lu_Wu:2011}.
Since the reassignment step in SST is based on the phase information, due to the mixed-up phase information in CWT, the reassignment result of CWT might deviate from the right location. Since there might be multiple components in the OAE signal, we consider the STFT-based SST to avoid this possible deterioration caused by the phase mixed-up.}} This is why we consider the STFT-based SST, particularly the second order SST. }

Here we suggest that it might be fairer to evaluate the performance by (i) comparing CWT with STFT-based SST {based on OTD}, 
and (ii) to use the \cite{LiuNeely-2010} kind of cochlear mechanics model to generate a signal and then evaluate performance by Loess smoothed group delay estimation. Results in Table \ref{tab:OTD} indicate that {ConceFT outperforms other methods, particularly when SNR is low, and this fits the theoretical development of ConceFT. Table \ref{tab:OTD} also demonstrates the potential of ConceFT as an alternative approach to handle other challenges for T-F analyses of TEOAE, including that TEOAEs are chirp-like, that there are multiple reflections within the cochlea, and that there could be SSOAEs because of multiple hot spots of generation. Nevertheless}, it does not seem that any of the afore-mentioned methods (CWT vs.~SST or ConceFT in particular) has absolutely better T-F analysis performance in all aspects, and {ConceFT provides a solution from a different angle in some situations. The answer to ``what is the best approach'' might depend on the application, and} whether the goal is to estimate the smoothed group delay or to follow the details of instantaneous frequency fluctuation. {A systematic study to answer this question, particularly for the real data, is needed and we expect to report our findings in the future work.}

\section{Conclusion}\label{sec:6}
{
Because of the random nature of TEOAE and existence of multiple reflections plus SSOAEs occasionally, we propose to model any given TEOAE signal as a sum of IMT functions in favor of flexibility of signal representation. Then, SST and MT
can be applied to obtained the ConceFT representation. ConceFT may have several advantages compared to commonly 
used and well-received methods in the OAE signal analysis field, such as CWT. First, it requires minimal prior assumptions about the underlying
signal, so it is less likely to lead to erroneous interpretation. Secondly, therefore, to extract underlying
information about individual components such as their IF and AM, one does not need to separate the components beforehand. Via analysis of simulated OAE-like signals under noisy conditions, we demonstrate that ConceFT indeed performs better than both the 1st-order and the 2nd-order SST, the CWT with a well-chosen mother wavelet, and two bilinear transforms,
in terms of its capability to preserve the ground truth.
Given the established rigor that supports the SST plus the noise robustness of conceFT thanks to MT, 
the proposed method has the potential to capture the time-varying IF function from individual TEOAEs reliably. 
A reasonable follow-up for this work would be to analyze real data in both normal ears and ears with
cochlea-related hearing impairment.
}

\section*{acknowledgments}
This research was supported by the Ministry of Science and Technology of Taiwan under grant No.~105-2628-E-007-005-MY2.

\appendix 

{
\section{Group delay in the mean sense}\label{sec:A}
In Eq.~(\ref{eq:1}), the function $\epsilon(x)$  is meant to characterize the unknown
perturbation of cochlear model parameters from smooth variation \citep{ZweigShera-1995}. 
Conceptually, the high frequency components in TEOAE
should appear prior to low frequency and this could possibly be shown via calculation of group delay as a function of frequency.
Here, we show that the group delay decreases against frequency 
\emph{in the mean sense}; that is, due to the variability among individuals, we regard the irregularity function $\epsilon$ as a random 
function among different ears which follows certain statistics. Under this randomness setup, the mean of the group delay across different ears decreases as $\omega$ increases \citep{SheraBergevin-2012}. 

A derivation is given as follows. Let $\Phi(\omega)$ denote the phase spectrum of $R(\omega)$; i.e., 
$R(\omega) = |R(\omega)|e^{i\Phi(\omega)}$. Since group delay involves calculating
the first derivative of $\Phi$ with respect to $\omega$, an intermediate mathematical step would be to take logarithm in the complex domain as to ``unraise'' $\Phi(\omega)$ from the exponent; that is, 
$\log\big(R(\omega)\big) = \log|R(\omega)| + i \Phi(\omega)$. Then, taking the first derivative with respect to $\omega$, we have
\begin{subequations}
\begin{eqnarray}
 &&\frac{1}{|R(\omega)|}\frac{\partial |R|}{\partial \omega} 
+i \frac{\partial \Phi(\omega)}{\partial \omega} \nonumber \\
&=& \frac{\partial \log R(\omega)}{\partial \omega}\label{eq:gd0}\\
	&=& \frac{1}{R(\omega)} \frac{\partial}{\partial\omega} \int \epsilon(x') e^{\frac{-(x'-x_p)^2}{2(\Delta x)^2}}e^{-i4\pi\left(\frac{x'-x_p}{\Lambda}\right)}dx' \label{eq:gd0.5}\\
	&=& \frac{1}{R(\omega)} \int \epsilon(x')\left[-\frac{(x'-x_p)}{(\Delta x)^2} + i\frac{4\pi}{\Lambda} \right]\frac{\partial x_p}{\partial\omega} \cdot e^{\frac{-(x'-x_p)^2}{2(\Delta x)^2}}e^{-i4\pi\left(\frac{x'-x_p}{\Lambda}\right)}dx'. \label{eq:gd1}
\end{eqnarray}
\end{subequations}
Note that Eq.~(\ref{eq:gd0.5}) is a simple application of the chain rule. Based on the log-linear relation in Eq.~(\ref{eq:x_p}), we have 
\[
\partial x_p/\partial\omega = -l/\omega. 
\]
Substituting this into Eq.~(\ref{eq:gd1}), the following expression is obtained,
\begin{subequations}
\begin{eqnarray}
\frac{\partial \log R(\omega)}{\partial \omega} &=&
		\frac{1}{R(\omega)}\frac{l}{\omega} \left[ -i \frac{4\pi}{\Lambda} R(\omega) + 
				\int \epsilon(x') H(x'-x_p,\omega)dx'\right] \\
		&=& -i \frac{4\pi}{\Lambda} \frac{l}{\omega} + \frac{1}{R(\omega)}\frac{l}{\omega}\int \epsilon(x') H(x'-x_p,\omega)dx', \label{eq:gd_final}
\end{eqnarray}
\end{subequations}
where $H(x,\omega)$ is defined as follows for the convenience of notations:
\[
	H(y,\omega) = \frac{y}{(\Delta x)^2} e^{\frac{-y^2}{2(\Delta x)^2}} e^{-i4\pi y/\Lambda}.
\]
Note that, in Eq.~(\ref{eq:gd_final}), the integral depends on $\epsilon(x')$ and generally does not vanish. Nevertheless,
if we treat $\epsilon$ as a random function and assume that  $\mathrm{E}\{\epsilon(x)\} = 0$ for all $x$, where $\mathrm{E}(\cdot)$ means to take the expected value across an ensemble of ears of similar conditions, then by comparing the imaginary and real parts of 
Eq.~(\ref{eq:gd_final}) and (\ref{eq:gd0}), we can conclude that the following relation holds for the group delay $\tau_g$ in the mean
sense,
\begin{equation}
	\mathrm{E}\{\tau_g(\omega)\} = \mathrm{E}\{-\frac{\partial \Phi}{\partial\omega}\} = \frac{4\pi}{\Lambda}\frac{l}{\omega} 
				= \frac{4\pi}{\Lambda}\left(-\frac{\partial x_p}{\partial\omega}\right)\,.
	\label{eq:Egd}
\end{equation}
Note that $\mathrm{E}\{\tau_g(\omega)\}$ monotonically decreases against $\omega$ in Eq.~(\ref{eq:Egd}). For
$\tau_g(\omega)$ in an individual ear, the relation should deviate from mean due to the presence of the integral term in Eq.~(\ref{eq:gd_final}).

When only the data from one single ear is available, the alternative way to estimate
the mean group delay is by smoothing over frequency. Interested readers can refer to
\cite{Keefe-2012} and \cite{SheraBergevin-2012} for a thorough evaluation of various smoothing methods.

\section{Formulation in the wavenumber domain}\label{sec:B}
In this subsection an interpretation of coherent reflection from the spatial frequency domain \cite{ZweigShera-1995} is studied. 
By transforming to the wavenumber domain and interchanging the order of integration, it turns out that
approximation can be made for the purpose of discussing time-domain properties of TEOAEs (to be revealed in Sec.~\ref{sec:C} and 
\ref{sec:D}). 

First, by defining $\rho(u) = e^{\frac{-u^2}{2(\Delta x)^2}} e^{-i4\pi\frac{u}{\Lambda}}$ and 
 applying a change-of-variable $y = x-x'$, Eq.~(\ref{eq:1}) can be re-written as,
\begin{equation}
	R(\omega) = \int \epsilon(y+x_p) \rho(y) dy.
\end{equation}
Now, let us assume that $\tilde{\epsilon}(k)$ is the spatial Fourier transform of $\epsilon(x)$ so we have
\begin{equation}
	\epsilon(y) = \frac{1}{2\pi}\int \tilde{\epsilon}(k) e^{iky} dk,
\end{equation}
where the variable $k$ is referred to as the wavenumber, or the spatial frequency. Combining the previous two equations, we have the following expression for $R(\omega)$,
\begin{subequations}
\begin{eqnarray}
R(\omega) &=& \frac{1}{2\pi}\int \Big(\int \tilde{\epsilon}(k) e^{ik(y+x_p)}dk\Big) \rho(y) dy \\
&=& \frac{1}{2\pi} \int \Big(\int \rho(y) e^{iky} dy\Big) e^{ikx_p}\tilde{\epsilon}(k) dk \\
&=& \frac{1}{2\pi} \int \tilde{\rho}^*(k) \tilde{\epsilon}(k) e^{ikx_p(\omega)} dk, \label{eq:7}
\end{eqnarray}
\end{subequations}
where $\tilde\rho(k) = \int \rho(y) e^{-iky} dy$ denotes the spatial Fourier transform of $\rho$, and $\tilde{\rho}^*(k) = \tilde{\rho}(-k)$.

Equation (\ref{eq:7}) has a \emph{spatial filtering} interpretation --- $\tilde\epsilon(k)$ is spatially filtered by $\tilde{\rho}^*(k)$, which has a peak magnitude at $k = 4\pi/\Lambda$ and a spatial bandwidth of $1/\Delta x$ \citep{ZweigShera-1995}. In this sense, we can say that as much as reflection is concerned, 
the most significant contribution stems from $\tilde{\rho}^*(k)\tilde{\epsilon}(k)$ at spatial frequency $k = 4\pi/\Lambda$.

\section{OAE evoked by a narrowband stimulus}\label{sec:C}
Now let us start from Eq.~(\ref{eq:7}) and discuss the special case of OAE evoked by a \emph{tone burst} (TB). 
Here, a tone burst $g(t)$ refers to a pure tone shaped in time by a window function so its spectrum is 
narrow-band around its center frequency $\omega_{b}$. 
TBs have long been used in human and animal experiments \citep{NeelyEA-1988, KonradMartinEA-2003, SiegelEA-2011} to study the properties of the cochlea.
Denoting the Fourier transform of $g(t)$ as $G(\omega)$, the first-reflection part of tone burst-evoked OAE (TBOAE) spectrum can be written as follows,
\begin{equation}
B(\omega) = R(\omega)G(\omega),
\label{eq:9}
\end{equation}
and thus its inverse Fourier transform $b(t)$ is
\begin{equation}
b(t) = \frac{1}{4\pi^2} \int \tilde\rho^*(k) \tilde{\epsilon}(k)  \Big(\int e^{ikx_p} G(\omega) e^{i\omega t} d\omega\Big) dk.
\label{eq:boft}
\end{equation}
Note that $x_p$ is a logarithmic function of $\omega$ so it seems that a close-form expression does not exist for
the integral $\int e^{ikx_p(\omega)} G(\omega) e^{i\omega t} d\omega$.
To continue, we utilize the assumption that $G(\omega)$ is narrow-banded. 
More precisely, we are going to assume that $\int e^{ikx_p(\omega)} G(\omega) e^{i\omega t} d\omega$ can be approximated by calculating within a narrowband 
$\omega \in [\omega_{b}-\delta, \omega_{b}+\delta]$.

Then, {by $x_p(\omega)=\log\frac{\omega_0}{\omega_{b}} - \log\frac{\omega}{\omega_{b}}$,} the following derivation can be made,
\begin{subequations}
\begin{eqnarray}
&& \int e^{ikx_p} G(\omega) e^{i\omega t} d\omega\nonumber\\
&\approx& \int_{\omega_{b}-\delta}^{\omega_{b}+\delta} 
e^{ikl\left(\log\frac{\omega_0}{\omega_{b}} - \log\frac{\omega}{\omega_{b}}\right)} G(\omega){e^{i\omega t}} d\omega   \\
&=& {C(k)} \int_{\omega_{b}-\delta}^{\omega_{b}+\delta} 
	e^{-ikl \log(1+u)} G(\omega) e^{i\omega t} d\omega \hspace{0.5cm}(\text{Here let } u=\frac{\omega}{\omega_b}-1)\\
&\approx& {C'(k)} \int_{\omega_{b}-\delta}^{\omega_{b}+\delta} e^{-ikl\frac{\omega}{\omega_b}}	e^{i\omega t} G(\omega) d\omega
																											\label{eq:taylor1}\\
&\approx& 2\pi {C'(k)} g\left(t-\frac{kl}{\omega_b}\right), \label{eq:approx1}
\end{eqnarray}
\end{subequations}
where ${C'(k)} = e^{ikl[1+\log(\omega_0/\omega_b)]}$ does not depend on $\omega$ and (\ref{eq:taylor1}) comes from Taylor's expansion around $\omega_b$, for $G(\omega)$ is assumed to be narrow-banded.

Substituting (\ref{eq:approx1}) into Eq.~(\ref{eq:boft}), we have the following approximation for $b(t)$, the first-reflection component of the TBOAE signal,
\begin{equation}
 b(t) \approx \frac{1}{2\pi} \int {C'(k)} \tilde{\rho}^*(k) \tilde{\epsilon}(k) g\left(t-\frac{kl}{\omega_b}\right) dk.
\label{eq:TB}
\end{equation}
Note that, if $\Delta x/\Lambda$ is sufficiently large, 
we can assume that $\tilde{\rho}^*(k)$ is concentrated around $k = 4\pi/\Lambda$ so the integral
in Eq.~(\ref{eq:TB}) can be regarded as mostly contributed by a short interval $\theta = [4\pi/\Lambda - \delta_k, 4\pi/\Lambda+ \delta_k]$, where $g(t-kl/\omega_b) \approx g(t-kl/\omega_b)|_{k=4\pi/\Lambda}$. Therefore,
 the following approximation for Eq.~(\ref{eq:TB}) is obtained,
\begin{equation}
 b(t) \approx  \Bigg(\int e^{ikl} e^{ikx_p(\omega_b)} \tilde{\rho}^*(k)\tilde{\epsilon}(k)dk\Bigg) \cdot
 g\left(t-\frac{4\pi}{\Lambda}\frac{l}{\omega_b}\right). 
\label{eq:btapprox}
\end{equation}
This expression shows some insights. First, the stimulus $g(t)$, which has a center frequency of $\omega_b$, is approximately
delayed by $4\pi l/(\omega_b\Lambda)$ when emitting out of the cochlea. 
This delay term agrees with our previous derivation in Eq.~(\ref{eq:Egd}) that was obtained from
a rather different angle. Secondly, 
assuming that the integral in Eq.~(\ref{eq:btapprox}) is again dominated by a short interval $\theta = [4\pi/\Lambda - \delta_k, 4\pi/\Lambda+ \delta_k]$, then $b(t)$ can be re-arranged as Eq.~(\ref{eq:14}).

\section{Breaking TEOAE into sub-band {TBOAE}}\label{sec:D}
The previous derivation assumes that the stimulus is narrow-band and the crude approximation in Eq.~(\ref{eq:taylor1}) 
assumes that higher order terms in the Taylor series expansion of $\log(1+u)$ can be omitted. In this section, we
attempt to loosen the narrow-band requirement on the stimulus.

The following Paley-Littlewood type decomposition is considered \cite{Stein:book}. 
Consider $\eta$ to be a smooth function that is compactly supported on $[-2,2]$ (Hz)\footnote{The exact unit here is not important for mathematical formulation, but to be consistent with notations, here we arbitrarily assign the unit Hz.} 
so that $\eta(x)=1$ when $x\in [-1,1]$ (Hz). Denote $\psi_j(x)=\eta(2^{-j}x)-\eta(2^{-j+1}x)$ for $j\in \mathbb{Z}$. $\psi_j$ could be viewed as a band-pass filter, which has the support over $[2^{j-1},2^{j+1}]$ (Hz). By a direct calculation, we see that $\eta(2^{-L_0+1}x)+\sum_{j=L_0}^\infty\psi_j(x)=1$ for all $x$, where $L_0\in \mathbb{Z}$.  
By the Paley-Littlewood type decomposition, the $R$ function in the Fourier domain can be rewritten as
\begin{equation}
R(\omega)=R(\omega)\Big[\eta(2^{-L_0+1}\omega)+\sum_{j=L_0}^{\infty}\psi_j(\omega)\Big]=R(\omega)\eta(2^{-L_0+1}\omega)+\sum_{j=L_0}^\infty R(\omega)\psi_j(\omega)\,,
\end{equation}
{ where $L_0$ is chosen to be a sufficiently low integer such that the support for $R(\omega)\eta(2^{-L_0+1}\omega)$ falls below
the human hearing range and the term can thus be ignored}. 
On the other hand, we model the incident wave to have a wide but compact support in the Fourier domain so that its Fourier transform is $\sum_{j\leq L_1}\psi_j$, where $L_1\geq 1$.
By these assumptions,  
the TEOAE signal have the following expansion:
\begin{equation}
P(t)=\int R(\omega)\sum_{j=L_0}^{L_1}\psi_j(\omega)e^{i\omega t}d\omega=\sum_{j=L_0}^{L_1}s_j\,.
\end{equation}
where in the time domain we have
\begin{equation}
s_j(t):=\mathcal{F}^{-1}[R(\omega)\psi_j(\omega)](t),
\end{equation}
which can be regarded as the $j$-th TBOAE with the dominant frequency around $2^j$ Hz.

Due to the log term appearing in the exponential, to better understand the TEOAE signal, we follow the approximation idea in Section \ref{sec:C}. 
Rewrite the $j$-th TBOAE as
\begin{equation}
s_j(t)=\frac{1}{2\pi}\int \tilde{\rho}^*(k)\tilde{\epsilon}(k)\Big[\int e^{i x_p(\omega)k}\psi_j(\omega)e^{i \omega t}d\omega\Big] dk\,,
\end{equation}
where the integration inside the bracket could be rewritten as
\begin{equation}
\int e^{i x_p(\omega)k}\psi_j(\omega)e^{i \omega t}d\omega=\int e^{-i  (kl \log\frac{\omega}{\omega_0}-t\omega)}\psi_j(\omega)d\omega\,.
\end{equation}
Recall that $\psi_j$ is supported on $[2^{j-1},\,2^{j+1}]$ Hz. By denoting $\Omega_j:=2^j$ and applying Taylor's expansion, we could approximate $kl \log\frac{\omega}{\omega_0}$ by $kl[ \log\frac{\Omega_j}{\omega_0} {-}\frac{1}{\Omega_j}(\Omega_j-\omega)-\frac{1}{{2}\tilde{\omega}^2}(\Omega_j-\omega)^2]$, where $\tilde{\omega}$ is between $\Omega_j$ and $\omega$. As a result, by ignoring the second order term, we have the following approximation
\begin{align}
\int e^{-i  (kl \log\frac{\omega}{\omega_0}-t\omega)}\psi_j(\omega)d\omega \approx&\, \int e^{i  (kl[ \log\frac{\Omega_j}{\omega_0}+\frac{1}{\Omega_j}(\Omega_j-\omega)]+t\omega)}\psi_j(\omega)d\omega\\
=&\,g_j\left(t-\frac{kl}{\Omega_j}\right)e^{i kl[\log\frac{\Omega_j}{\omega_0}+1]}\,,\nonumber
\end{align}
where $g_j(t):=\mathcal{F}^{-1}(\psi_j)(t)$, and hence approximate $s_j$ by
\begin{equation}
s^{(L)}_j(t):=\frac{1}{2\pi}\int \tilde{\rho}^*(k)\tilde{\epsilon}(k)     g_j\left(t-\frac{kl}{\Omega_j}\right)e^{ikl[\log\frac{\Omega_j}{\omega_0}+1]}      dk\,.
\end{equation}
{By assumption, $\tilde{\rho}^*(k)$ decays exponentially fast and is concentrated on $\frac{4\pi}{\Lambda}$ and $\tilde{\epsilon}(k) e^{i kl[\log\frac{\Omega_j}{\omega_0}+1]}$ is bounded. Therefore, by a direct approximation, $s^{(L)}_j(t)$ becomes
\begin{align}
s^{(L)}_j(t)\approx&\,\frac{1}{2\pi}\left[\int \tilde{\rho}^*(k)\tilde{\epsilon}(k)    e^{ikl\log\frac{\Omega_j}{\omega_0}}      dk\right]  g_j\left(t-\frac{4\pi l}{\Lambda\Omega_j}\right) e^{i \frac{4\pi}{\Lambda}l}  
\label{eq:22}\\
=&\,\frac{1}{2\pi}R(\Omega_j) 2^jg_1\left (2^jt-\frac{4\pi l}{\Lambda}\right)e^{i \frac{4\pi}{\Lambda}l} 
\nonumber
\end{align}
where} we use the fact that 
\begin{equation}
g_j(t)=\frac{1}{2\pi}\int^{2^{j+1}}_{2^{j-1}}[\eta(2^{-j}\omega)-\eta(2^{-j+1}\omega)]e^{i \omega t}d\omega=\frac{2^j}{2\pi}\int^2_{1/2}\psi_1(\omega)e^{-i 2^j\omega t}d\omega=2^jg_1(2^jt)\,,
\end{equation}
which is an oscillatory signal (or could be understood as a dilated wavelet).
This approximation suggests that we could well approximate a TBOAE as a ``time lagged'' reflected signal, where the reflected signal comes from an inner ear {that has a locally linearized tonotopic mapping relation}.
As a result, we have the following approximation {
\begin{equation}\label{FinalExpansion:TEOAE}
P(t)\approx \frac{1}{2\pi}e^{i \frac{4\pi}{\Lambda}l}  \sum_{j= L_0}^{L_1}  R(\Omega_j) 2^jg_1\left (2^jt-\frac{4\pi l}{\Lambda}\right)\,,
\end{equation}
}where we view the TEOAE as a summation of a sequence of latent TBOAE of different frequencies, and the latency depends on the frequency of the TBOAE.
Note that the TEOAE has a higher frequency oscillation in the beginning with a stronger power and a short period, and has a lower frequency oscillation later with a weaker power and a longer period. {The relationship between the frequency and latency is inverse to each other, and the decay of the amplitude depends on the decay of $R$.} To sum up, this model depicts the fundamental feature of first-reflection component of TEOAE  --- as an oscillatory signal, and both the amplitude and the frequency decrease as time goes.

\end{document}